\documentclass{article}

\usepackage[utf8]{inputenc}
\usepackage[dvipdfmx]{graphicx}
\usepackage{siunitx}
\usepackage{authblk}
\usepackage{hyperref}
\usepackage{amsmath}
\usepackage{bm}

\usepackage{cancel}
\usepackage{amsmath,amssymb}
\usepackage{hyperref}

\usepackage{ulem}
\usepackage{upgreek}
\usepackage[none]{hyphenat}
\usepackage{gensymb}

\usepackage{siunitx}
\usepackage{authblk}
\usepackage{bm}

\usepackage{cancel}
\usepackage{amsmath,amssymb}
\usepackage{hyperref}
\usepackage{here}

\usepackage{color}
\usepackage{here}
\numberwithin{equation}{section}

\usepackage{ulem}

\usepackage{amsthm}
\newtheoremstyle{query}%
{}{}
{\color{red}}
{}
{\sffamily\bfseries}{:}{12pt}
{}
\theoremstyle{query}
\newtheorem{aq}{Author Query/Comment}

\newcommand{\baq}{\begin{aq}}
\newcommand{\eaq}{\end{aq}}

\usepackage{lineno}

\hypersetup{
    colorlinks=true,
    linkcolor=blue,
    filecolor=magenta,      
    urlcolor=cyan,
    }

\usepackage[margin=2.5cm]{geometry}

\title{\vspace{-10mm} Background-oriented Schlieren technique with vector tomography for measurement of axisymmetric pressure fields of laser-induced underwater shock waves}

\author[1]{\small{Sayaka Ichihara}}
\author[1]{Takaaki Shimazaki}
\author[1,2]{Yoshiyuki Tagawa}

\affil[1]{Department of Mechanical Systems Engineering, Tokyo University of Agriculture and Technology, Koganei Campus 6-204, 2-24-16 Nakacho, Koganei, Tokyo, Japan
}
\affil[2]{Institute of Global Innovation Research, Tokyo University of Agriculture and Technology, Koganei Campus 6-204, 2-24-16 Nakacho, Koganei, Tokyo, Japan
}

\date{}

\begin{document}


\maketitle

\vspace{-10mm}

\paragraph{Abstract} 
This study aims to overcome the problems that existing background-oriented schlieren (BOS) techniques based on computed tomography  (CT-BOS) face when  measuring pressure fields of laser-induced underwater shock waves. To do this, it proposes a novel BOS technique based on vector tomography (VT-BOS) of an axisymmetric target.  
The remarkable feature of the proposed technique is the reconstruction of an axisymmetric vector field with nonzero divergence, such as the field of a laser-induced underwater shock wave. 
This approach is based on an approximate relation between the projection of the axisymmetric vector field and the reconstructed vector field.
For comparison, the pressure fields of underwater shock waves are measured with VT-BOS, CT-BOS, and a needle hydrophone.
It is found that VT-BOS is significantly better than CT-BOS in terms of better convergence, less dependence on the spatial resolution of the acquired images, and lower computational cost.
The proposed technique can be applied  not only to fluid dynamical fields, but also to other axisymmetric targets in other areas, such as electromagnetics and thermodynamics.
\vspace{20pt}

\section{Introduction}
Laser-induced underwater shock waves\cite{bell1967laser,lauterborn2013shock} are widely used in clinical medicine  for various techniques, such as needle-free injection \cite{tagawa2013needle,kiyama2019visualization,krizek2020needle,miyazaki2021dynamic,hayasaka2017effects}, lithotripsy \cite{razvi1996intracorporeal,sankin2008focusing}, skull base tumor removal \cite{ogawa2011pulsed}, and treatment of diseases such as femoral head osteonecrosis \cite{ludwig2001high,wang2015extracorporeal,hausdorf2010shock}, osteochondritis dissecans \cite{heidersdorf2000osteochondritis}, osteochondral lesions (\cite{zhang2020extracorporeal}), and bone marrow edema \cite{cao2021bone}.
These techniques have the advantage that  incisions are not necessary, and  thus there is a lower risk of complications developing \cite{shrivastava2005shock,wang2015extracorporeal}.
It is therefore of great value to investigate underwater shock waves in the context of  medical applications.
Previous studies have used point measurement systems such as  hydrophones, but these disturb the flow field or impose restrictions on the experimental conditions \cite{herbert2006cavitation}.
Therefore, noncontact pressure measurements are required to enable further developments in this field. 

Existing noncontact pressure measurement methods include pressure-sensitive paint (PSP) \cite{liu1997temperature,bell2001surface,mitsuo2003temperature}, particle image velocimetry (PIV)\cite{van2013piv,fujisawa2006pressure,elsinga2006tomographic}, and background-oriented schlieren (BOS)\cite{yamamoto2015application,hayasaka2016optical} techniques.
The BOS technique has the advantage of allowing measurement with simple equipment such as a background image and a camera \cite{meier1998new,venkatakrishnan2004density,hayasaka2019mobile}.
It has been used to measure a variety of density fields, for example, those in flames \cite{grauer2018instantaneous}, supersonic
flows \cite{ota2011computed,ota2015quantitative}, and surface dielectric barrier discharges \cite{kaneko2021background}.
The prominent feature of the BOS technique  is that it can be used  for noncontact measurements of the pressure field of underwater shock waves \cite{hayasaka2016optical}, and, as described below, it has great scope for further development in this context.

\begin{figure}[H]
\begin{center}
\centering
\includegraphics[width=0.8\columnwidth]{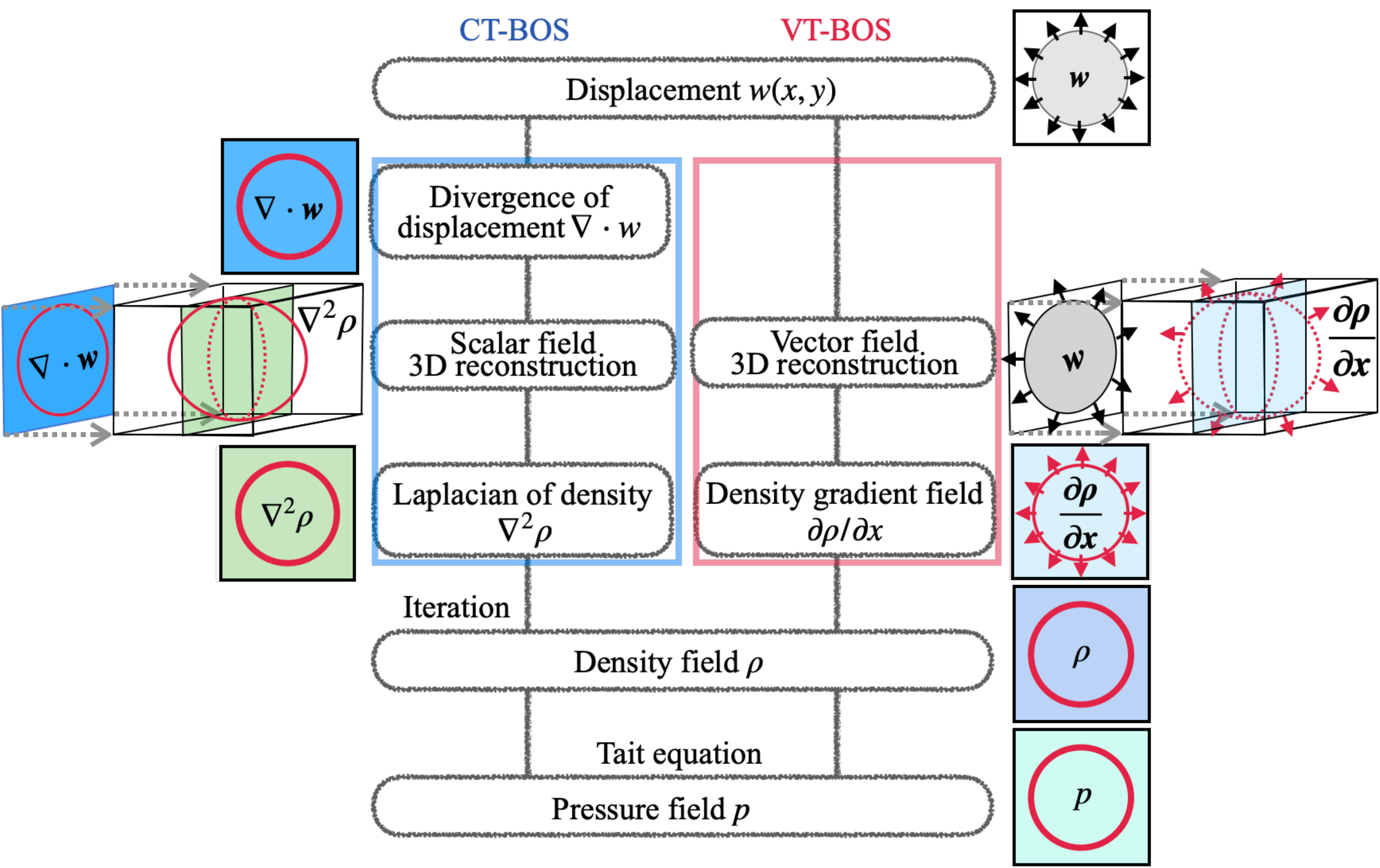}
\caption{Calculation procedures for CT-BOS (the previous technique) and VT-BOS (the proposed technique). }
\label{fig:bos}
\end{center}
\end{figure}

The procedure of the BOS technique for calculating the density and pressure of a target is shown in Fig.\ref{fig:bos}.
The apparent displacement is calculated by comparing  background images with and without the measurement target present.
The calculated apparent displacement is proportional to the integral of the density gradient with respect to the optical axis of the camera.
Therefore, it is necessary to calculate the three-dimensional (3D) density gradient field from the apparent displacement by 3D reconstruction.

In conventional 3D reconstruction, instead of the vector field of apparent displacement $\boldsymbol{w}$, reconstruction is performed on a scalar field, namely, the divergence $\nabla \cdot \boldsymbol{w}$ of $\boldsymbol{w}$ \cite{venkatakrishnan2005density,atcheson2008time}. 
Existing BOS techniques use 3D reconstruction methods, such as filtered back-projection (FBP)\cite{venkatakrishnan2005density,tipnis2013density,sourgen2012reconstruction} and the algebraic reconstruction technique (ART)\cite{atcheson2008time,leopold2013reconstruction,ota2011computed}, which are types of computed tomography (CT).
The BOS technique with computed tomography is hereinafter referred to as CT-BOS. 
After CT calculation in CT-BOS, the Laplacian 
 $ \nabla^2 \rho$ ($\propto \nabla \cdot \boldsymbol{w}$) of the density field $\rho$ is obtained and used in the Poisson equation for $\rho$, which is then solved  iteratively.
From the linear relationship between density and pressure, the pressure field is calculated \cite{hayasaka2016optical}.
When the Poisson equation is solved from the 3D reconstruction of a scalar field by iteration in CT-BOS,  problems arise with regard, for example, to the accuracy of pressure calculation and convergence of the solution \cite{venkatakrishnan2005density,hayasaka2016optical}, the dependence on resolution \cite{hayasaka2016optical,yamamoto2015application}, and the high computational cost \cite{nicolas2016direct,ihrke2004image}.

To solve these problems, we apply vector tomography (VT), which can directly reconstruct a 3D vector field (hereinafter referred to as the reconstructed distribution) from apparent displacements (hereinafter referred to as the projected distribution), in the BOS technique.
VT is a reconstruction method defined by Norton\cite{norton1989tomographic} and Prince\cite{prince1996convolution,prince1994tomographic}.
This method extends the central-slice theorem \cite{natterer2001mathematical} used in 3D reconstruction of scalar field and realizes the reconstruction of a vector field when the divergence  of the reconstructed distribution is zero \cite{norton1989tomographic,svetov2014numerical}.
VT is used, for example, to reconstruct 3D magnetization vector fields \cite{donnelly2018tomographic}, coronal magnetic fields \cite{kramar2013vector}, and  electrostatic potential fields \cite{wolf2019holographic}.
However, the reconstructed distribution of vector fields such as laser-induced shock waves measured in this paper does not have a divergence  of zero, making it difficult to apply conventional VT to such fields.

Here we  derive an approximate matrix equation relating the reconstructed and projected distributions by assuming that the reconstructed distribution is an axisymmetric vector field with only radial components.
Based on this equation, we propose a novel VT that can perform reconstruction without the need to satisfy the above conditions {\cite{Ichihara2022}} .
It is possible to reconstruct the density gradient field of an axisymmetric shock wave \cite{tagawa2016pressure} with respect to the optical axis of a laser pulse.

By applying this VT method to the BOS technique, the density gradient field can be obtained from the displacement, which is a vector field.
Integration is performed on the density gradient field to obtain the density field.
The  pressure field, which has a linear relationship with the density field, can then be calculated.
This new BOS technique using VT is hereinafter referred to VT-BOS.
We believe that VT-BOS can avoid the calculation procedures that cause problems in CT-BOS and can overcome these problems.

The remainder of this paper is organized as follows. The theoretical basis of the proposed VT-BOS method is described in Sec.~\ref{sec:2}.
Experimental measurements of underwater shock waves are presented in Sec.~\ref{sec:3_exp}).
Using these experimental data and data from previous studies \cite{hayasaka2016optical}, we investigate the validity of VT-BOS by comparing the accuracy of the calculated pressure, the convergence of the calculations, the dependence on spatial resolution, and the computational cost of VT-BOS with those of CT-BOS in Sec.~\ref{sec:4_result_discussion}. Our conclusions are presented in Sec.~\ref{sec:5}

\section{\label{sec:2}Theory of BOS technique}
\subsection{\label{sec:2_bos}Background-oriented schlieren technique}
\begin{figure}[ht]
    \centering
    \includegraphics[width=0.8\columnwidth]{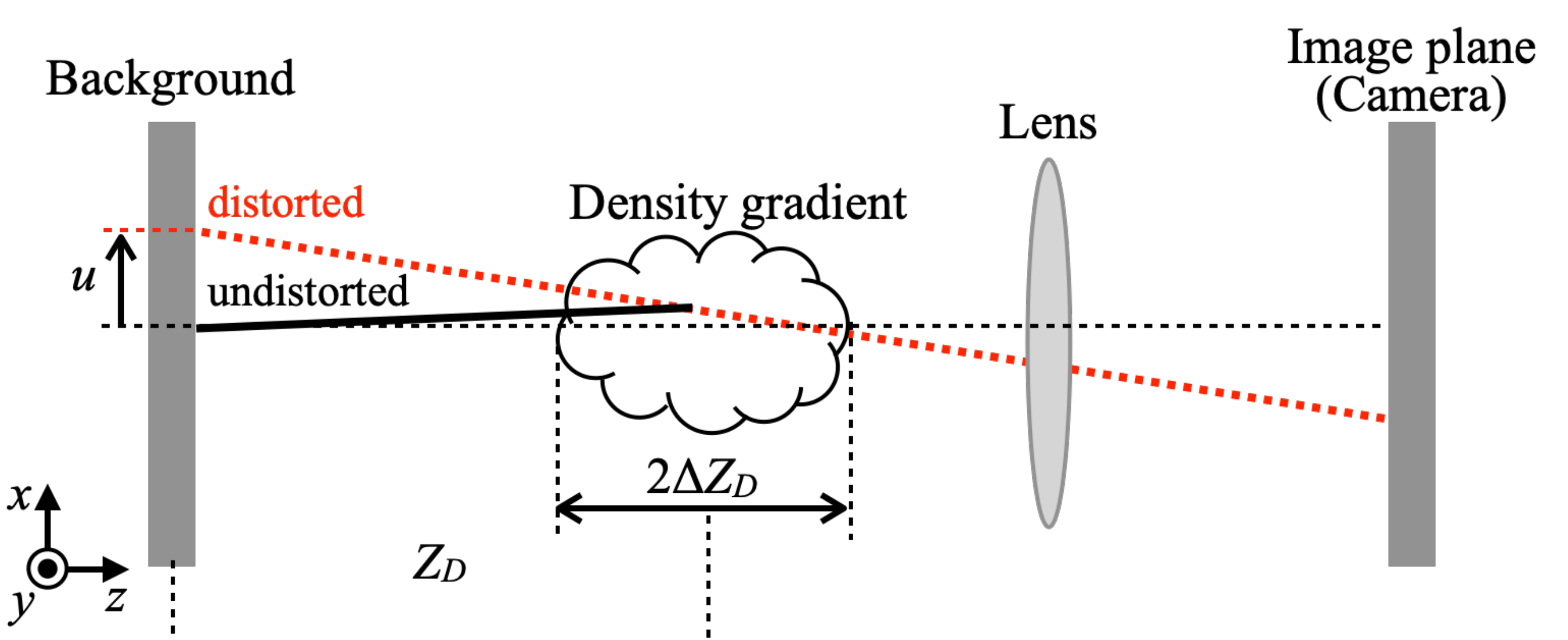}
    \caption{Principle of BOS technique.}
    \label{fig:theory_bos}
\end{figure}
Fig.~\ref{fig:theory_bos} shows a conceptual diagram of the BOS technique \cite{meier2002computerized,venkatakrishnan2004density,yamamoto2015application}.
In the BOS method, the camera, the measurement target, and the background are placed on a straight line.
The camera takes the background image with and without the measurement target.
The apparent local displacement vector $(u,v)$ on the image is obtained from the two images using an image analysis technique such as digital image correlation (DIC) 
\cite{meunier2003analysis,moisy2009synthetic}, 
optical flow \cite{atcheson2009evaluation,brox2004high}, or
fast checkerboard demodulation (FCD) 
\cite{hatanaka2013background,wildeman2018real,shimazaki2022}. 
An $(x,y,z)$ coordinate system is adopted, with the background image in the $xy$ plane and the 
camera's line of sight in the $z$ direction (Fig.~\ref{fig:theory_bos}).
The relationship between the apparent displacement $u$ in the $x$ direction and the refractive index gradient is given by 
\begin{equation}
u = \frac{Z_D}{n_0}\int^{Z_D + \Delta Z_D}_{Z_D - \Delta Z_D} \frac{\partial n}{\partial x}\,dz,
\label{eq:u}
\end{equation}
\cite{venkatakrishnan2004density}, 
where $n$ is the refractive index of the object being measured, $n_0$ is the refractive index of the surrounding fluid, $Z_D$ is the distance from the background to the object being measured, and $\Delta Z_D$ is the half-width of the object.
The Gladstone--Dale equation \cite{merzkirch2012flow, raffel2015background} holds between the refractive index $n$ and the density $\rho$:
\begin{equation}
n = \rho K + 1,
\label{eq:grandstone_dale}
\end{equation}
where $K$ ($=3.14 \times 10^{-4}$~m$^3$/kg) 
is the Gladstone--Dale constant. 
Substituting Eq.~\ref{eq:grandstone_dale} into Eq.~\ref{eq:u}, the equation relating the $x$-direction displacement to the density gradient is obtained as
%
\begin{equation}
u = \frac{Z_D K}{n_0}\int^{Z_D + \Delta Z_D}_{Z_D - \Delta Z_D} \frac{\partial \rho}{\partial x}\,dz.
\label{eq:u-rho}
\end{equation}
Similarly, the equation relating the $y$-direction displacement $v$ to the density gradient is
\begin{equation}
v = \frac{Z_D K}{n_0}\int^{Z_D + \Delta Z_D}_{Z_D - \Delta Z_D} \frac{\partial \rho}{\partial y}\,dz.
\label{eq:v-rho}
\end{equation}
There are two methods for obtaining the density from $u$ and $v$: vector field 3D reconstruction (vector tomography, VT) and scalar field 3D reconstruction (computed tomography, CT), which are described in Secs.~\ref{sec:2_VT} and \ref{sec:2_CT}, respectively.

Using the calculated density, the pressure is calculated by Tait's formula \cite{brujan2010cavitation,richardson1947hydrodynamic}
\begin{equation}
\frac{p + B}{p_0 + B} = \left( \frac{\rho}{\rho_0} \right)^\alpha,
\label{eq:pressure}
\end{equation}
where $p_0$ is the atmospheric pressure, $\rho_0$ ($= 998$~kg/m$^3$) is the density of the ambient fluid (in this case the density of standard-state water), and $B$ ($= 314$~MPa) and $\alpha$ ($=7$) are constants \cite{brujan2010cavitation}.

\subsection{\label{sec:2_VT}Density field calculation using vector tomography}
\begin{figure}[H]
\centering
\includegraphics[width=0.8\columnwidth]{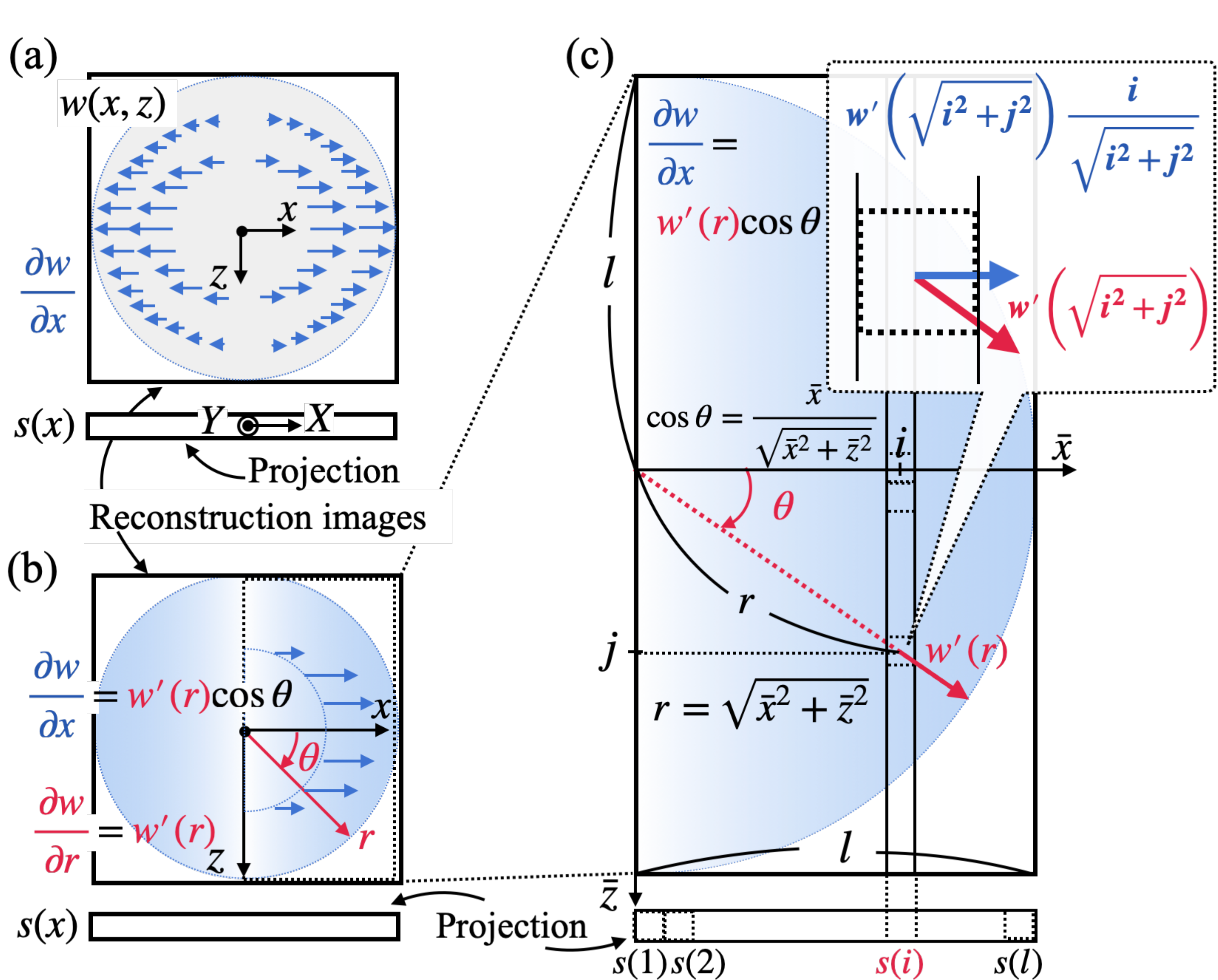}
\caption{(a) and (b) Projected values on the $X$ axis and reconstructed distribution of a vector field in  $(x,z)$ coordinates. The blue vectors and blue zone in (b) show the reconstructed distribution of the vector field $\partial w / \partial x$. The gray zone in (a) shows the distribution of the scalar field $w$. The red vectors in (b) show the reconstructed distribution of the vector field $\partial w / \partial r$ in $(r, \theta)$ coordinates. (c) Enlarged cross section from the reconstructed distribution and projected values in (b), where $x \ge 0$ and $X \ge 0$.}
\label{fig:recon_vr1}
\end{figure}
\begin{figure}[H]
\centering
\includegraphics[width=0.8\columnwidth]{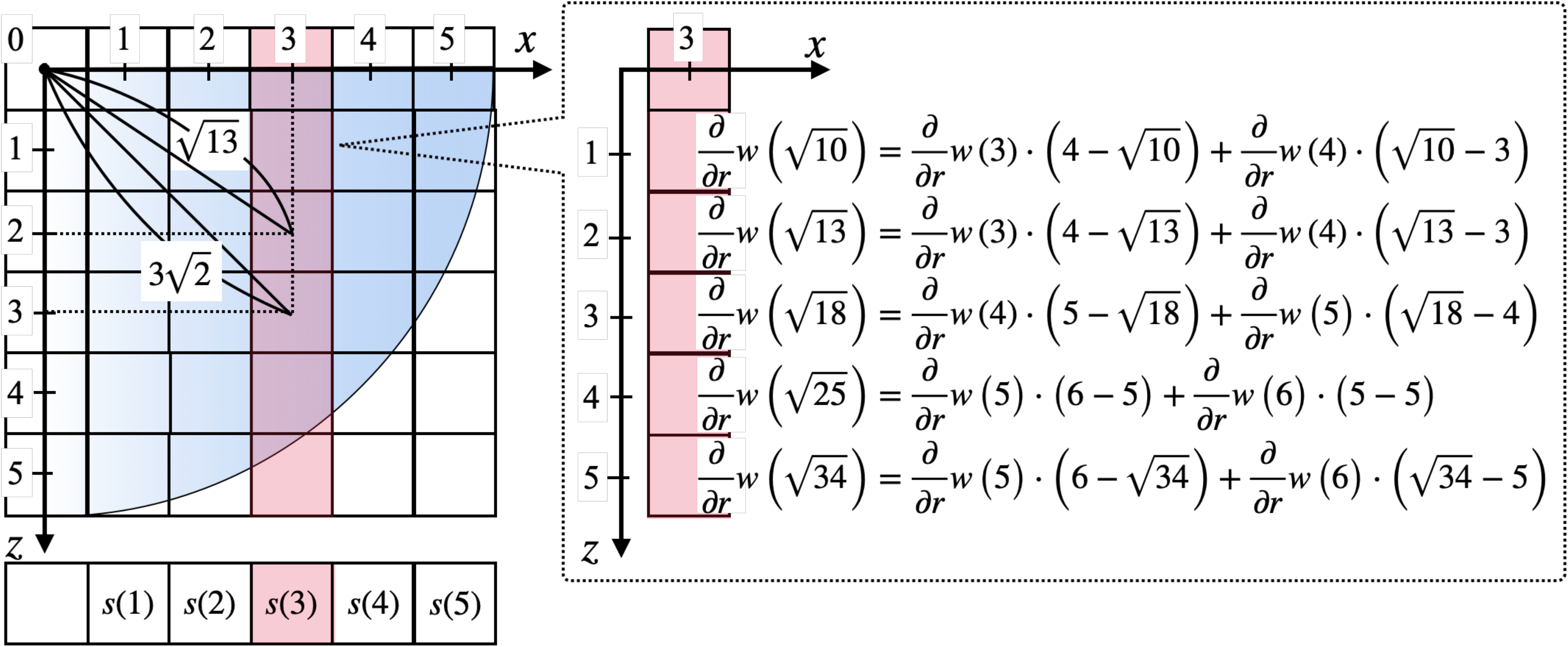}
\caption{Relationship between reconstructed distribution and projected values in a $5\times5$ matrix. 
The enlargement shows specific formulas for linear interpolation on the $x=3$ axis.}
\label{fig:recon_vr2}
\end{figure}

In this section, we describe a new VT method for computing density fields.
Three coordinate systems are used: $(X,Y)$, $(x,y,z)$, and $(r,\theta)$.
First, we explain the relationship between these coordinate systems and the properties of the target vector field.

As shown in Fig.~\ref{fig:recon_vr1}(a), the 2D vector field projected from the 3D vector field is represented in $(X,Y)$ coordinates, and the 3D reconstructed distribution vector field is represented in $(x,y,z)$ coordinates.
The $X$ and $x$ axes and the $Y$ and $y$ axes are parallel, and the origin of the $(X,Y)$ coordinates is on the $z$ axis.
The vector field in $(X,Y)$ coordinates is the projection of the reconstructed distribution in $(x,y,z)$ coordinates onto the $z$ axis.
The reconstructed distribution is assumed to be $y$-axis symmetric.
Therefore, the projected vector field is $Y$-axis symmetric.
Furthermore, the reconstructed distributions of all $xz$ sections are assumed to be vector fields with $z$-axis symmetry and only a radial component.
The radial component is the $r$-direction component in $(r,\theta)$ coordinates, which are the polar coordinates corresponding to the $(x,z)$ Cartesian coordinates (see Fig. \ref{fig:recon_vr1}(b)).
Based on the above assumptions, the reconstructed distribution on the $xz$ cross section is not related to the $Y$ component of the projected distribution, which is a vector field.
Therefore, the VT proposed in this paper focuses only on the $X$ component of the projected vector field.

Let the reconstructed distribution (vector field) be represented by a potential (scalar field) $w(x,z)$ (the gray area in Fig.~\ref{fig:recon_vr1}(a)) distributed on the $xz$ plane.
That is, the $x$ component of the reconstructed distribution is represented by   the partial derivative of $w(x,z)$ with respect to $x$, that is, $\partial w / \partial x$.
Furthermore, let the $\partial w / \partial x$ distribution be the reconstructed distribution on the $xz$ plane and $\bm{s}(X)$ be the projected field of $\partial w / \partial x$ along the $z$ axis.
Since $\partial w / \partial x$ has no $Y$ component, neither does the projected field $\bm{s}(X)$.
Assuming that $X=x$, the relationship between the reconstructed distribution and the projected values is
\begin{equation}
\bm{s}(x) = \int \frac{\partial w}{\partial x}\,  dz.
\label{eq:vr_int}
\end{equation}
From the characteristics of the reconstructed distribution, $\partial w / \partial x$ is a vector field symmetric about the origin  and  about the $z$ axis on the $xz$ plane (see Figs.~\ref{fig:recon_vr1}(a) and (b)).
Also, $\partial w/ \partial x$ can be expressed in terms of the radial vector $\partial w / \partial r$ in the 2D $(r,\theta)$ polar coordinate system  and the angle $\theta$ with the $x$ axis (Fig.~\ref{fig:recon_vr1}(b)): $\partial w/ \partial x = (\partial w / \partial r ) \cos \theta$.
Here, $w(r,\theta)$, which is a polar transformation of a scalar field $w(x,z)$, has no $\theta$ component, and so $\partial w / \partial \theta$ is zero.
Therefore, hereinafter, $\partial w(r,\theta)/ \partial r$ will be written as $\partial w(r)/ \partial r$.
In terms of the Cartesian coordinate system,  $r$ can be expressed as $r= \sqrt{x^2+z^2}$ and $\theta$ as $\cos \theta = x/ \sqrt{x^2+z^2}$, and so Eq.~\ref{eq:vr_int} can be rewritten as
\begin{equation}
\bm{s}(x) = \int \frac{\partial }{\partial r} w\left( \sqrt{x^2+z^2} \right) \frac{x}{\sqrt{x^2+z^2}}\, dz.
\label{eq:vr_int3}
\end{equation}

As shown in Fig.~\ref{fig:recon_vr1}(b), $\partial w/ \partial x$ is distributed symmetrically about the origin and along the $z$ axis, and so we focus only on $xz$ sections where $x$ is greater than or equal to 0, as shown in Fig.~\ref{fig:recon_vr1}(c).
We then consider the final discretization of Eq.~\ref{eq:vr_int3} and express it in terms of an approximate matrix relationship
\begin{equation}
\begin{bmatrix}
\bm{s}(1)\\
\bm{s}(2)\\
\vdots\\
\bm{s}(\bar{r})\\
\vdots\\
\bm{s}(N)\\
\end{bmatrix}
\propto
\begin{bmatrix}
\dfrac{\partial}{\partial r} w(1)\\[9pt]
\dfrac{\partial}{\partial r} w(2)\\
\vdots\\
\dfrac{\partial}{\partial r} w(\bar{r})\\
\vdots\\
\dfrac{\partial}{\partial r} w(N)\\
\end{bmatrix}
.
\label{eq:vr_gyo_propto}
\end{equation}
Note that $\bar{r}$ and $N$ are natural numbers greater than 1, with  $1 \le \bar{r} \le N$.
First, Eq.~\ref{eq:vr_int3} is rewritten in discrete instead of continuous form: 
\begin{equation}
\bm{s}(\bar{x}) =  \frac{\partial}{\partial r}w
\left( \bar{x} \right) 
+ 2\sum_{\bar{z}=1}^{N} 
\frac{\partial}{\partial r}w
\left( \sqrt{\bar{x}^2+\bar{z}^2} \right) \frac{\bar{x}}{\sqrt{\bar{x}^2+\bar{z}^2}}.
\label{eq:vr_sig1}
\end{equation}
Note that $\bar{x}$ and $\bar{z}$ are natural numbers, and the 
 coordinates $(\bar{x},\bar{z})$ are discrete coordinates as shown in Fig.~\ref{fig:recon_vr2}.
The relation between the discretized coordinates $(\bar{x},\bar{z})$ and the original coordinates $(x,z)$ is
\begin{equation}
    \begin{split}
     x = c \bar{x}, \\
     z = c \bar{z},
    \end{split}
\label{eq:coor}
\end{equation}
where $c$ [pixel$^{-1}$] is a constant.

At a certain position $\bar{x}=\bar{x}_0 $, Eq.~\ref{eq:vr_sig1}
can be written as
\begin{equation}
\bm{s}(\bar{x}_0 )= 
\frac{\partial}{\partial r}w(\bar{x}_0 )+
2\sum_{\bar{z}=1}^{N} 
\frac{\partial}{\partial r}w
\left( \sqrt{\bar{x}_0 ^2+\bar{z}^2} \right) \frac{\bar{x}_0 }{\sqrt{\bar{x}_0 ^2+\bar{z}^2}}.
\label{eq:vr_sig3}
\end{equation}
In Eq.~\ref{eq:vr_sig3}, $\sqrt{\bar{x}_0 ^2+\bar{z}^2}$ in the term $(\partial/\partial r) w(\sqrt{\bar{x}_0 ^2+\bar{z}^2})$ is not necessarily natural numbers.
On the other hand, in the term $(\partial/\partial r) w(\bar{r})$ of the reconstructed distribution in Eq.~\ref{eq:vr_gyo_propto}, the variable $\bar{r}$ ($=1,2,\dots,N$) is a natural number.
When Eq.~\ref{eq:vr_sig3} is written in a discrete matrix form such as Eq.~\ref{eq:vr_gyo_propto}, the real number $(\partial/\partial r)w
\left( \sqrt{\bar{x}_0 ^2+\bar{z}^2} \right)$ in Eq.~\ref{eq:vr_sig3} must be expressed using $(\partial/\partial r)w
( \bar{r} )$ ($\bar{r}=1,2,\dots,N$).
By linear interpolation, the real number $(\partial/\partial r)w
\left( \sqrt{\bar{x}_0 ^2+\bar{z}^2} \right)$ can be written as 
\begin{equation}
\begin{split}
\frac{\partial}{\partial r}w
\left( \sqrt{\bar{x}_0 ^2+\bar{z}^2} \right)=
\frac{\partial}{\partial r} w \left( b_{\bar{x}_0 ,\bar{z}} \right)
\left( b_{\bar{x}_0 ,\bar{z}} + 1 - \sqrt{\bar{x}_0 ^2+\bar{z}^2} \right)\\
+\frac{\partial}{\partial r} w \left( b_{\bar{x}_0 ,\bar{z}}+1 \right) 
\left( \sqrt{\bar{x}_0 ^2+\bar{z}^2} - b_{\bar{x}_0 ,\bar{z}} \right),
\label{eq:vr_interp}
\end{split}
\end{equation}
where $b_{\bar{x}_0 ,\bar{z}}$ is a natural number satisfying $b_{\bar{x}_0 ,\bar{z}} \le \sqrt{\bar{x}_0 ^2+\bar{z}^2} < b_{\bar{x}_0 ,\bar{z}} + 1 $.
When the linear interpolation in Eq.~\ref{eq:vr_interp} is applied to Eq.~\ref{eq:vr_sig3}, the latter becomes 
\begin{equation}
\bm{s}(\bar{x}_0 )=\frac{\partial}{\partial r}w(\bar{x}_0 )
+2\sum_{\bar{z}=1}^{N} 
\left[
\frac{\partial}{\partial r} w \left( b_{\bar{x}_0 ,\bar{z}} \right)
\left( b_{\bar{x}_0 ,\bar{z}} + 1 - \sqrt{\bar{x}_0 ^2+\bar{z}^2} \right)
 +  \frac{\partial}{\partial r} w \left( b_{\bar{x}_0 ,\bar{z}}+1 \right) 
\left( \sqrt{\bar{x}_0 ^2+\bar{z}^2} - b_{\bar{x}_0 ,\bar{z}} \right)
\right]
\label{eq:vr_sig_interp}
\end{equation}
If the coefficients of $(\partial/\partial r) w ( b_{\bar{x}_0 ,\bar{z}} )$ are taken to be constants $\alpha_{\bar{x}_0 ,\bar{z}}$, 
Eq.~\ref{eq:vr_sig_interp} becomes 
\begin{equation}
\bm{s}(\bar{x}_0 ) = 
\sum_{\bar{z}=1}^{N} 
 \alpha_{\bar{x}_0 ,\bar{z}} \frac{\partial}{\partial r} w(b_{\bar{x}_0 ,\bar{z}}).
\label{eq:vr_sig_inp2}
\end{equation}
From Eq.~\ref{eq:vr_sig_interp}, at some position $\bar{x} = \bar{x}_0 $, the term $ b_{\bar{x}_0 ,\bar{z}}$ in the reconstructed distribution   satisfies $\bar{x}_0  \le b_{\bar{x}_0 ,\bar{z}} \le N$, and so the terms $(\partial/\partial r) w( 1 )$ to $(\partial/\partial r) w( \bar{x}_0 -1 )$ are not included in the summation.
Therefore, the coefficients $\alpha_{\bar{x}_0 ,\bar{z}}$ of the  terms $(\partial/\partial r)w ( b_{\bar{x}_0 ,\bar{z}} )$ in the reconstructed distribution are 0 for $1 \le \bar{z} \le \bar{x}_0 -1$.
Also, the coefficients $\alpha_{\bar{x}_0,\bar{z}}$ are  determined only by the position in the discrete coordinates $(\bar{x},\bar{z})$.

In the projection $\bm{s}(\bar{x})$ in Eq.~\ref{eq:vr_sig_inp2},  the variable $\bar{x}$ takes values among all the natural numbers $1,2, \dots $, and we can then show that
\begin{equation}
\bm{s}(\bar{x}) = 
\sum_{\bar{z}=1}^{N} 
 \alpha_{\bar{x},\bar{z}} \frac{\partial}{\partial r} w(b_{\bar{x},\bar{z}}), 
\label{eq:vr_sig_inp3}
\end{equation}
which can be written in matrix form as
\begin{subequations}
\label{eq:vr_gyo_last}
\begin{align}
\bm{S}&=\bm{A}\bm{W},\label{eq:vr_gyo_last-a}\\
\intertext{or, explicitly,} 
\begin{bmatrix}
\bm{s}(1)\\
\bm{s}(2)\\
\bm{s}(3)\\
\vdots\\
\bm{s}(n)
\end{bmatrix}
&=
\begin{bmatrix}
\alpha_{1,1}&\alpha_{1,2}&\alpha_{1,3}&\cdots &\alpha_{1,N}\\
0&\alpha_{2,2}&\alpha_{2,3}&\cdots&\alpha_{2,N}\\
0&0&\alpha_{3,3}&\cdots&\alpha_{3,N}\\
\vdots&\vdots&\vdots&\ddots&\vdots\\
0&0&0&\cdots&\alpha_{N,N}
\end{bmatrix}
\begin{bmatrix}
\dfrac{\partial}{\partial r} w(1)\\[9pt]
\dfrac{\partial}{\partial r} w(2)\\[9pt]
\dfrac{\partial}{\partial r} w(3)\\[9pt]
\vdots\\
\dfrac{\partial}{\partial r} w(N)
\end{bmatrix}
.
\label{eq:vr_gyo_last-b}
\end{align}
\end{subequations}
As mentioned earlier, the coefficients $\alpha_{\bar{x}_0 ,\bar{z}}$ in Eq.~\ref{eq:vr_sig_inp2} for the reconstructed distribution   at some location $\bar{x} = \bar{x}_0 $ are zero for $1 \le \bar{z} \le \bar{x}_0 -1$.
Similarly, the coefficient $\alpha_{2,\bar{z}}$ in the reconstructed distribution at $\bar{x} = 2$ is 0 for $\bar{z} = 1$, and the coefficient $\alpha_{3,\bar{z}}$ of the reconstructed distribution at $\bar{x} = 3$ is 0 for $\bar{z} = 1,2$.
It can be seen that the coefficient matrix $\bm{A}$ in Eq.~\ref{eq:vr_gyo_last} for the reconstructed distribution for all natural numbers $\bar{x}$ ($\bar{x}\ge 1$) is always an upper triangular matrix with nonzero and nonsingular diagonal components.
It therefore  has an inverse matrix $\bm{A}^{-1}=\bm{\widetilde{A}}/|\bm{A}|$, where $\bm{\widetilde{A}}$ and $|\bm{A}|$ are the  respectively adjugate matrix and determinant of $\bm{A}$  \cite{harville1998matrix}.
Multiplication of  both sides of Eq.~\ref{eq:vr_gyo_last-a} by the inverse matrix $\bm{A}^{-1}$ then gives
\begin{equation}
\bm{A}^{-1}\bm{S} = 
\bm{A}^{-1} \bm{A W} 
 =\bm{W}
.   
\label{eq:vr_gyo2}
\end{equation}
Thus, the reconstructed distribution in matrix form $\bm{W}$ can be calculated by simply multiplying the matrix of projected values $\bm{S}$ by the inverse  of the coefficient matrix $\bm{A}$,
and the integral relation in Eq.~\ref{eq:vr_int} between the projected values of the vector field and the reconstructed distribution of the vector field has been transformed into the matrix form in Eq.~\ref{eq:vr_gyo2} by VT.

The integral equation~\ref{eq:u-rho} shows the relationship between the projected value of the vector field and the reconstructed distribution of the vector field in the BOS technique.
When applying VT to Eq.~\ref{eq:u-rho}, the displacement $u$ in Eq.~\ref{eq:u-rho} corresponds to  the projected vector field $\bm{s}(x)$ in Eq.~\ref{eq:vr_int}, and the density gradient field $\partial \rho(x,y) /\partial x$ in Eq.~\ref{eq:u-rho} corresponds to the reconstructed distribution $\partial w/\partial x$ in Eq.~\ref{eq:vr_int}.
The displacement $u$ and density gradient $\partial \rho(x,y)/\partial x$ are represented in a discretized coordinate system.
Applying VT using Eq.~\ref{eq:vr_int} to the displacement $\bm{u}(\bar{x},\bar{y_0})$ and density gradient $\partial \rho ( \bar{r} ) /\partial \bar{r}$ ($= \partial \rho (\bar{x},\bar{y_0})/ \partial \bar{x}$) at the position $\bar{y}=\bar{y_0}$ in Eq.~\ref{eq:u-rho} gives
\begin{equation}
\frac{\partial \rho (\bar{r})}{\partial r}= 
\frac{n_0}{Z_DK}\bm{A}^{-1}\bm{u}( \bar{x}, \bar{y_0}).
\label{eq:vr_bos}
\end{equation}
Note that $\bar{r}$ here denotes the radial component of the discretized polar coordinates.
By applying VT to all the $y$ displacements $\bm{u}(\bar{x},\bar{y})$ on the $y$ axis, the density gradient distribution $ \partial \rho (\bar{x},\bar{y})/ \partial x$ in the $xy$ section at $z=0$ can be calculated.
The density field $\rho$ is calculated by integrating in the $x$ direction over the calculated density gradient field $\partial \rho (x,y)/ \partial x $:
\begin{equation}
\rho = \int^{\infty}_{0} \frac{ \partial \rho (x,y) }{ \partial x} \,dx
\label{eq:vr_integral}
\end{equation}
Here, the density of the surrounding fluid $\rho_0$ is taken to be that of the fluid outside  the measurement target (i.e., the shock wave).

\subsection{\label{sec:2_CT}Density field calculation using computed tomography }

In CT, reconstruction is performed from the divergence  of the projected field \cite{venkatakrishnan2005density,atcheson2008time,sourgen2012reconstruction,leopold2013reconstruction,hayasaka2016optical}.
In this paper, we use the analytical method of filter back-projection (FBP) \cite{konkani2013springer,ziegler2007noise} for CT, following  Hayasaka \cite{hayasaka2016optical}. 
The underwater shock wave is $y$-axis symmetric \cite{sankin2008focusing,tagawa2016pressure} because it was generated by pulsed laser irradiation in the $y$-axis direction in the coordinate system shown in Fig.~\ref{fig:theory_bos}.
Therefore, the projection onto the plane perpendicular to the $z$ axis is assumed to be the same from all directions, and reconstruction is performed using only images taken from one direction, as shown in Fig.~\ref{fig:theory_bos}.
Taking the divergences of the displacements  $u$ (Eq.~\ref{eq:u-rho}) and $v$ (Eq.~\ref{eq:v-rho}) and applying FBP, we obtain  the Poisson equation for the density on the plane containing the axis of symmetry (the $y$ axis) \cite{hayasaka2016optical} (Fig.~\ref{fig:bos}): 
\begin{equation}
\frac{\partial^2 \rho}{\partial x^2} + 
\frac{\partial^2 \rho}{\partial y^2} =
-\frac{2 (1 + K \rho_0)}{\Delta Z_D K(\Delta Z_D + 2 Z_D)} 
\left( 
\frac{\partial U}{\partial x} + 
\frac{\partial V}{\partial y}
\right),
\label{eq:poisson}
\end{equation}
Let $U$ and $V$ be the displacements of the reconstructed distribution in the $x$ and $y$ directions, respectively.
The FBP calculation uses the iradon function implemented in MATLAB.
The density field is calculated by solving the Poisson equation~\ref{eq:poisson} iteratively  (through successive over-relaxation, SOR) (Fig.~\ref{fig:bos}).
The convergence condition for this iterative procedure is defined as 
\begin{equation}
E(n) = \frac{\rho^{n}_{i,j} - \rho^{n-1}_{i,j}}{\rho^{n}_\mathrm{max}}
\le E_0,
\label{eq:convergence}
\end{equation}
where $n$ is the iteration step and $(i,j)$ is the position in $(x,z)$ coordinates.
 $\rho_\mathrm{max}^n$ is the maximum value of the density at the $n$th iteration step.
When $E(n)$ is less than a certain value (the convergence value $E_0$) at all positions, the iterative computation is said to have converged.
The determination of  the convergence value is discussed in Sec.~\ref{sec:4_pre}.

\section{\label{sec:3_exp}Experiments}
\subsection{Experimental setup}
\begin{figure}[h]
    \centering
    \includegraphics[width=0.8\columnwidth]{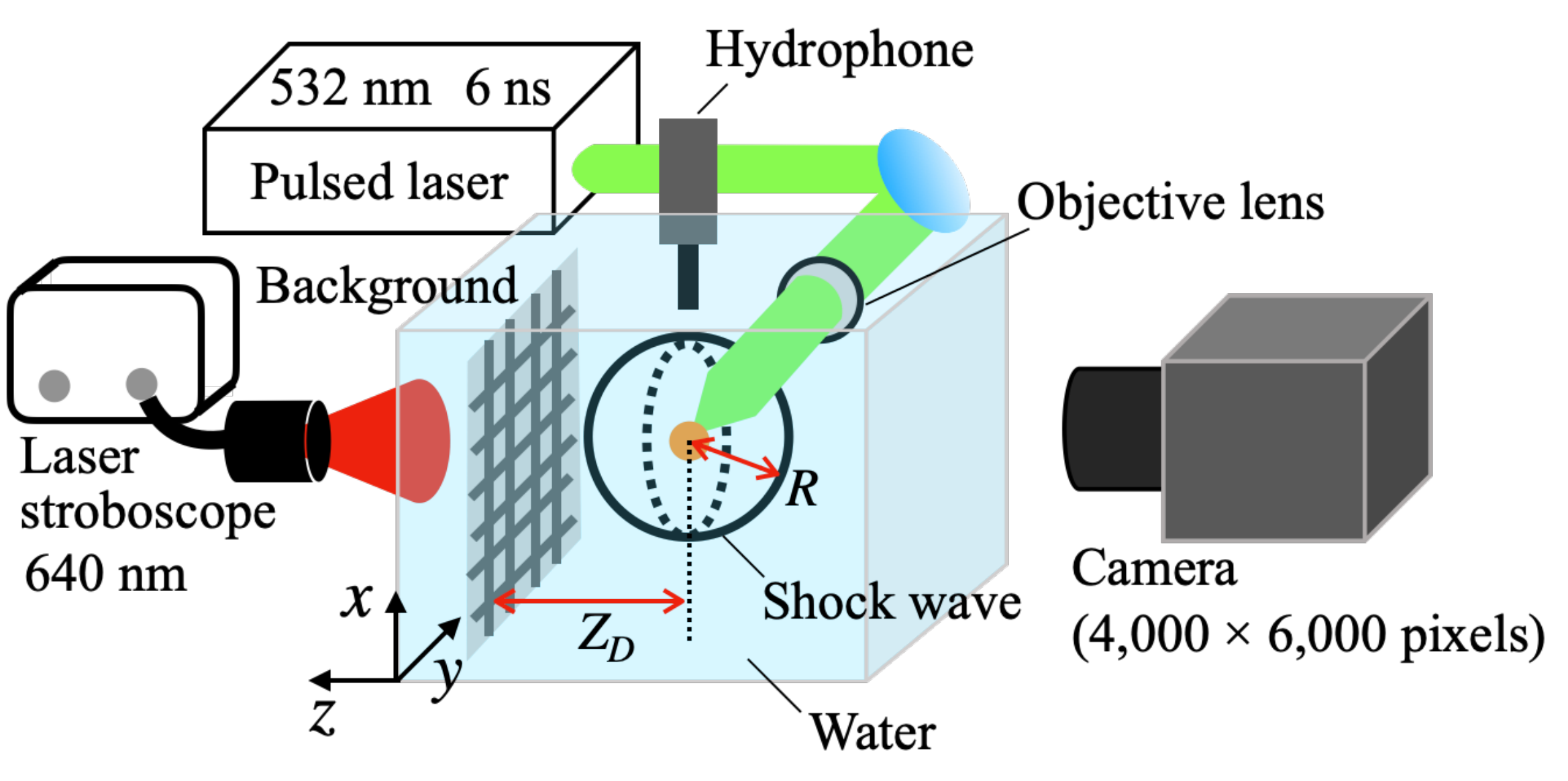}
    \caption{Experimental setup.}
    \label{fig:exp}
\end{figure}

The experimental methods follow those of \cite{hayasaka2016optical} and \cite{shimazaki2022}.
The camera, target, and background are placed on the same line, as shown in Fig.~\ref{fig:exp}.
The grid or checker pattern background (grid width  20--40~\si{\um}) is placed perpendicular to the camera's line of sight.
The object to be measured (i.e., an underwater shock wave) is generated between the camera and the background.
The distance from the center of the shock wave to the background is $Z_D$.
The underwater shock wave to be measured is a field with large local displacement, and 
if the displacement field is too large, its detection may become difficult.
Since the displacement field increases in proportion to $Z_D$ under given measurement conditions (see Eq. \ref{eq:u-rho}), 
$Z_D$ should be about 1.5--2.5~mm to ensure that the  displacement field does not become too large.
A pulsed laser  for illumination (SI-LUX640, Specialised Imaging, wavelength  640~nm, pulse width  20~ns) is placed behind the background image.
An Nd:YAG pulsed laser  (Nano S PIV, Litron Lasers
Ltd., wavelength  532~nm, pulse width  6~ns) is focused into the water by an objective lens (SLMPLN20X, Olympus), 
generating shock waves from the focal point.
A high-spatial-resolution camera (EOS 80D, Canon, spatial resolution $4000 \times 6000$ pixels) is used to take a background image (Fig.~\ref{fig:disp_field}(a)) in the absence of a shock wave and a background image (Fig.~\ref{fig:disp_field}(b)) when a shock wave is generated (shutter speed  0.5~s).
The pulsed lasers for generating the shock wave and  for illuminating the background image are both connected to a delay generator (Model 575, BNC), and the timing of shooting is adjusted so that the range of the shock wave radius $R$ is 2.0--3.5~mm.
\cite{hayasaka2016optical} have pointed out that high-resolution images are necessary for sufficient accuracy of pressure calculation, 
and therefore the resolution of the captured images should be 0.9--1.4~\si{\um}/pixel.
A hydrophone (Muller-Platte Needle Probe, Mueller Instruments, effective radius $<$0.25~mm) is placed on the shock wave center axis perpendicular to the camera's line of sight, about 0.2--0.3 mm from the shock wave radius, to measure the shock wave pressure.
In this experiment, the shock wave pressure peak $P_\mathrm{max}$ is measured under three different conditions.
We also use part of the data from \cite{hayasaka2016optical} as reference data (see Table~\ref{tab:exp_cond}).
To simplify the analysis, the image is rotated 90$^{\circ}$ along the $y$ axis (Figs.~\ref{fig:disp_field}(a) and \ref{fig:disp_field}(b)), as shown in Fig.~\ref{fig:exp}.
Two 90$^{\circ}$-rotated images (Figs.~\ref{fig:disp_field}(a) and \ref{fig:disp_field}(b)) are analyzed.

\begin{table*}[ht]
 \centering
\caption{Experimental conditions.}
\begin{tabular}{l c c c c} 
Dataset& i  & ii& iii&iv\\ 
&  & & &\cite{hayasaka2016optical}\\ \hline \hline 
$p_\mathrm{max}$ [MPa] & 0.4 & 1.0 &2.4 & 1.2 \\
$Z_{D}$ [mm] & 1.93 &2.01&1.50 & 8.00 \\
$R$ [mm] & 2.1 & 2.3 & 3.5& 5.0  \\
Spatial resolution [\si{\um}/pixel] & 1.16 &0.91& 1.31 &5.08 \\
Image resolution [pixels]& $2535\times1499$ &$2947\times1885$& $2941\times1796$ &$1979\times1046$ \\
\end{tabular}
\label{tab:exp_cond}
\end{table*}

\subsection{Pressure calculation method}
\begin{figure}[H]
\centering
\includegraphics[width=0.8\columnwidth]{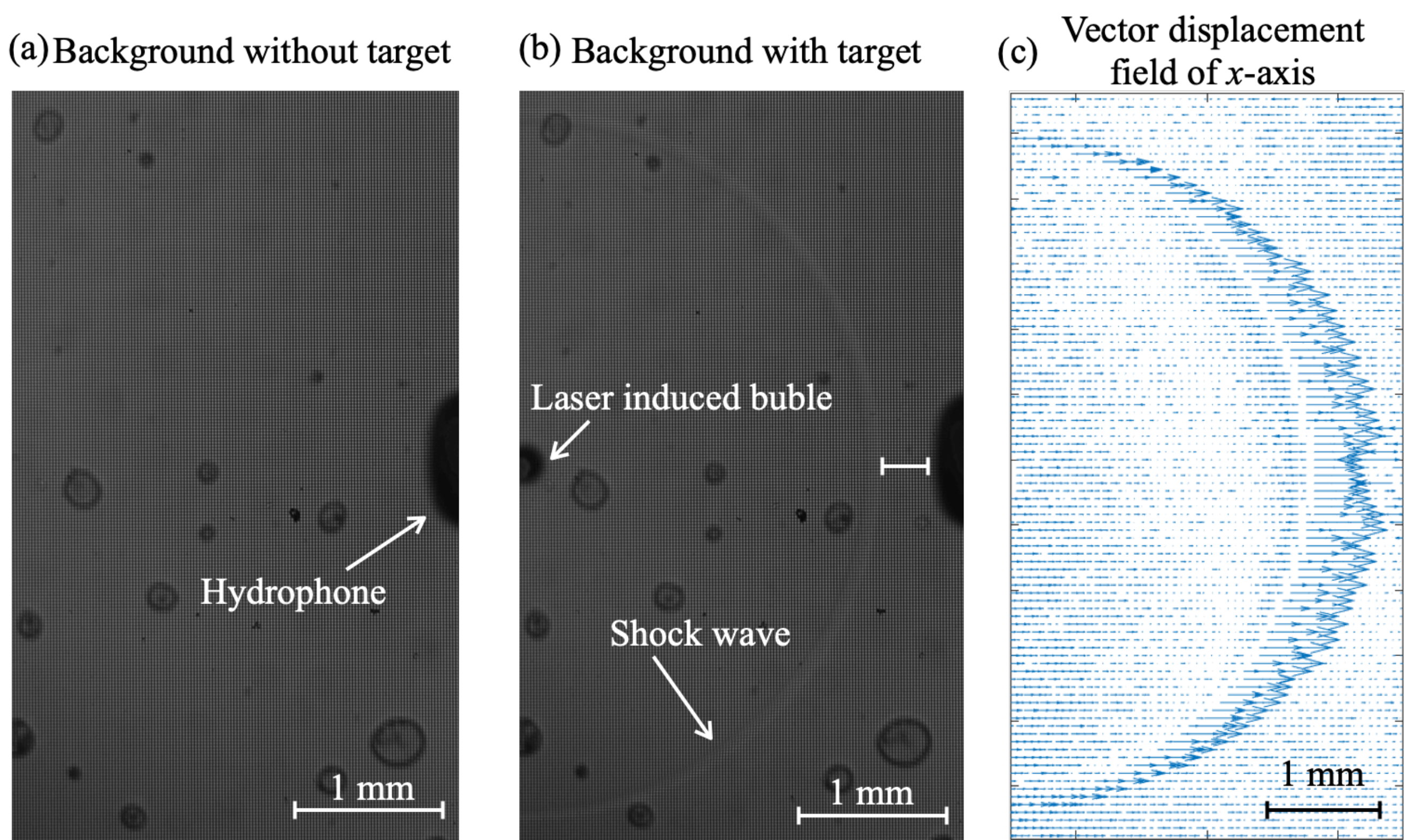}
\caption{(a) and (b) Captured background images respectively with and without the target (shock wave). (c) Vector field of $x$ component  measured by displacement detection (FCD).}
\label{fig:disp_field}
\end{figure}

\begin{figure}[H]
\centering
\includegraphics[width=0.8\columnwidth]{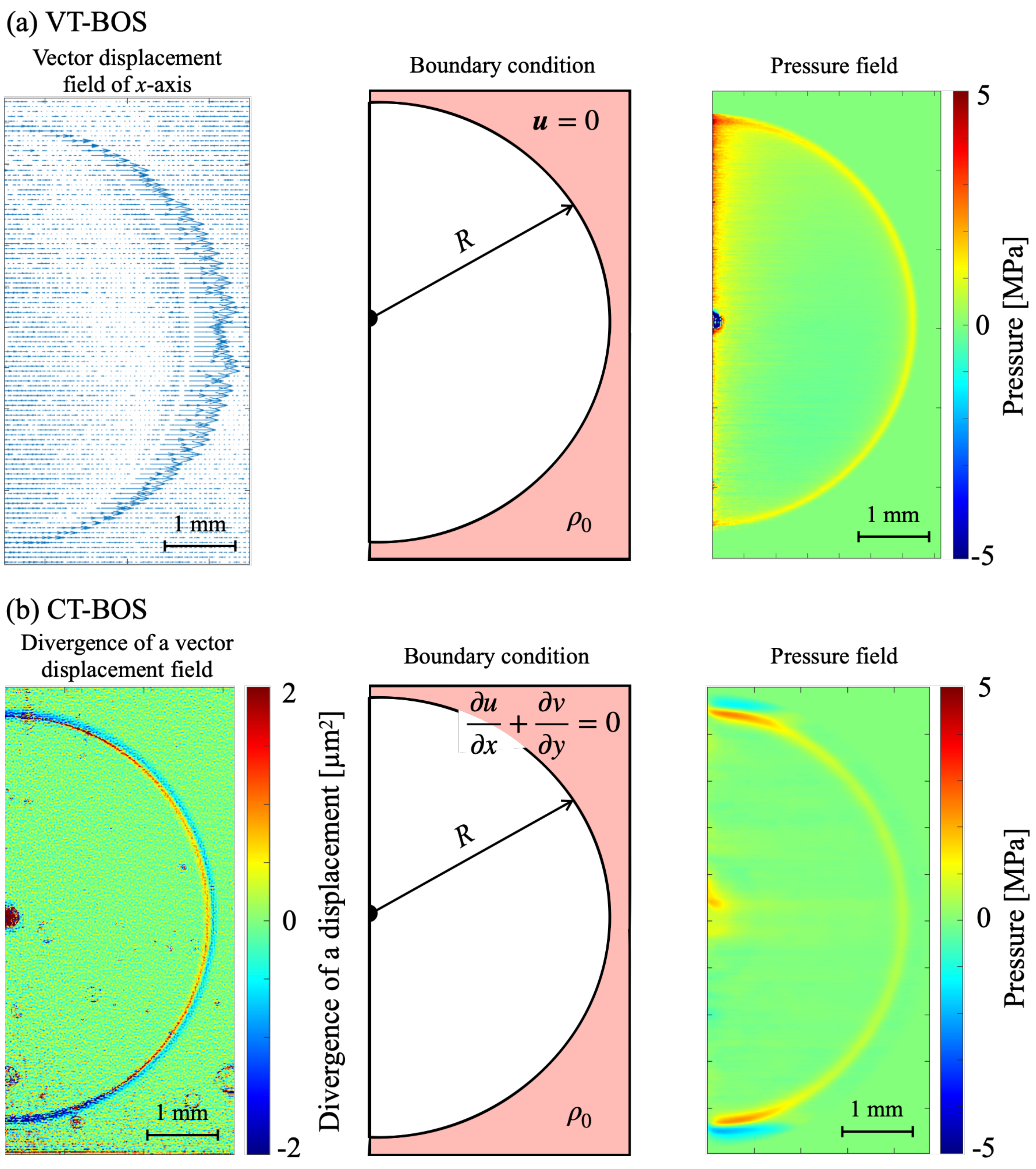}
\caption{(a) Left: Vector field of $x$ component (blue vectors). Center: Boundary conditions for VT-BOS. Right: Pressure field measured by VT-BOS. (b) Left: Divergence of displacement. Center: Boundary conditions for CT-BOS. Right: Pressure field measured by CT-BOS.}
\label{fig:cond}
\end{figure}

In this subsection, we describe the procedure for analyzing the BOS. Fast checkerboard demodulation 
(FCD) (\cite{wildeman2018real,shimazaki2022}) is applied to the two background images taken in the experiment (Figs.~\ref{fig:disp_field}(a) and \ref{fig:disp_field}(b))  to obtain the $x$-direction displacement vector field of the shock wave, as shown in Fig.~\ref{fig:disp_field}(c).
In VT-BOS, VT is applied to the displacement vector field in the $x$ direction (Fig.~\ref{fig:cond}(a), left) to calculate the density gradient field.
Outside  the displacement and shock wave (Fig.~\ref{fig:cond}(a), center, in red), as boundary conditions, the displacement field $\bm{u}$ is taken to be zero and the density is taken to be that of water at 25\,$^\circ$C under atmospheric pressure, $\rho_0$.
This region outside  the shock wave is defined as the region outside a circle of radius $R+50$ pixels, where $R$ is the radius of the shock wave.
The density and pressure fields (Fig.~\ref{fig:cond}(a), right) are calculated by integrating the density gradient field.
In CT-BOS, CT (Sec.\ref{sec:2_CT}) is applied to the divergence of the displacement vector field (a scalar field) (Fig.~\ref{fig:cond}(b), left).
Outside  the displacement and shock wave (Fig.~\ref{fig:cond}(b), center, in red), as the boundary condition,  the divergence  of the displacement $\partial u/\partial x +\partial v/\partial y$ is taken to be zero.
In CT-BOS, after reconstruction, the density field is calculated by solving the Poisson equation~\ref{eq:poisson} from Sec.\ref{sec:2_CT} using the SOR method.
In the iterative calculation, the portion of the laser-induced bubble at the center of the shock wave is taken to have a pressure of zero.
The error that defines the convergence condition~\ref{eq:convergence} is also calculated at the same time.
After the density field has been calculated, the pressure field is calculated using Tait's equation~\ref{eq:pressure}.  
The pressures of the shock wave calculated by both techniques are compared with the pressure measured by the hydrophone.

\subsection{Evaluation methods}
\begin{figure}[H]
\centering
\includegraphics[width=0.8\columnwidth]{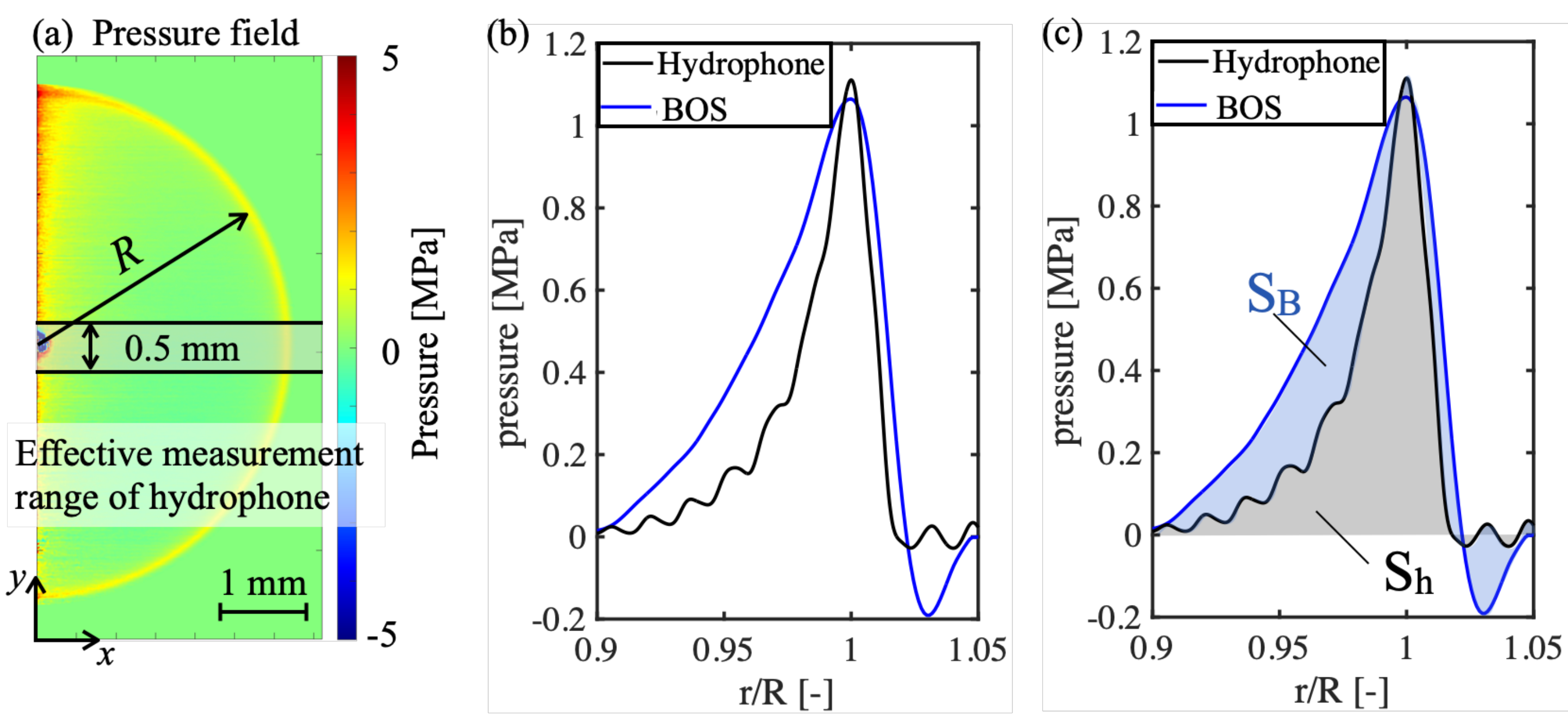}
\caption{(a) Effective measurement range of the  hydrophone  compared with the pressure field measured by BOS techniques.
(b) Comparison of pressures measured by BOS and by a hydrophone on the $x$ axis. 
The horizontal axis shows the distance from the center of the shock wave, nondimensionalized by the radius $R$ of the shock wave.
(c) The blue area $S_B$ represents the difference between the pressure measured by the BOS technique and that measured by the hydrophone, and the gray area $S_h$ represents the  the pressure measured by the hydrophone.}
\label{fig:pre_compare}
\end{figure}
\subsubsection{\label{sec:3_ratio}Method of comparison of pressure calculation accuracy and convergence of solutions}
The accuracies of CT-BOS and VT-BOS pressure calculations are compared based on the pressure measured by the hydrophone.
Unlike the BOS technique, the hydrophone measures shock wave pressure at a small measurement range. 
Therefore, in the pressure field of the shock wave calculated by the BOS technique, the pressure field in the effective measurement range of the hydrophone is averaged in the $y$-axis direction (the shaded area in Fig.~\ref{fig:pre_compare}(a)).
Then, as shown in Fig.~\ref{fig:pre_compare}(b), the pressure calculated by the BOS technique is compared with the hydrophone result.
The closer the pressure value from the BOS technique is to the hydrophone measurement, the greater the measurement accuracy of the BOS technique.

For a  quantitative comparison of the accuracy of pressure calculations, we proceed as follows.
We calculate the area $S_B$  between the  pressure curve calculated by the BOS method and the measured pressure curve  (Fig.~\ref{fig:pre_compare}(c), blue area) and 
 the area $S_h$  between the measured pressure curve and the zero-pressure line (Fig.~\ref{fig:cond}(d), gray area). We then define the ratio $S=(S_B+S_h)/S_h$. The closer this ratio is to 1, the closer is the hydrophone pressure measurement  to the pressure calculated by  the BOS method and thus the more accurate is the latter calculation.

Investigation of convergence  for CT-BOS  requires an iterative approach.
A solution is said to be convergent if the convergence condition~\ref{eq:convergence} is satisfied at some iteration.
VT-BOS does not require iteration, and the density gradient can be uniquely calculated from the displacements using a matrix equation.

\subsubsection{\label{sec:3_sp}Reduction of spatial resolution}
To investigate the relationship between spatial resolution and the accuracy of pressure calculation by CT-BOS and VT-BOS, the spatial resolution of the original image is reduced and the results of pressure calculations by CT-BOS and VT-BOS are compared.
CT-BOS changes the spatial resolution of the divergence  of the displacement field after reconstruction, whereas VT-BOS changes the spatial resolution of the displacement field itself.
The spatial resolution of the image with reduced resolution is given by that of the original image (0.91--5.08 \si{\um}m/pixel) multiplied by a factor $\eta$. Specifically, the spatial resolution is reduced using the interp2 function of MATLAB.
Table~\ref{tab:exp_cond} shows the image resolution and spatial resolution for the different experimental conditions.


\subsubsection{\label{sec:3_cputime}
Calculation  of computational cost}
The computational costs are compared using MATLAB's CPUtime function.
This function returns the total CPU time, allowing comparison of computational cost independently of PC performance.



\section{\label{sec:4_result_discussion}Results and discussion}
\subsection{\label{sec:4_pre}Pressure calculation accuracy and convergence}
\subsubsection{Pressure calculation accuracy}
\begin{figure}[H]
\centering
\includegraphics[width=1.0\columnwidth]{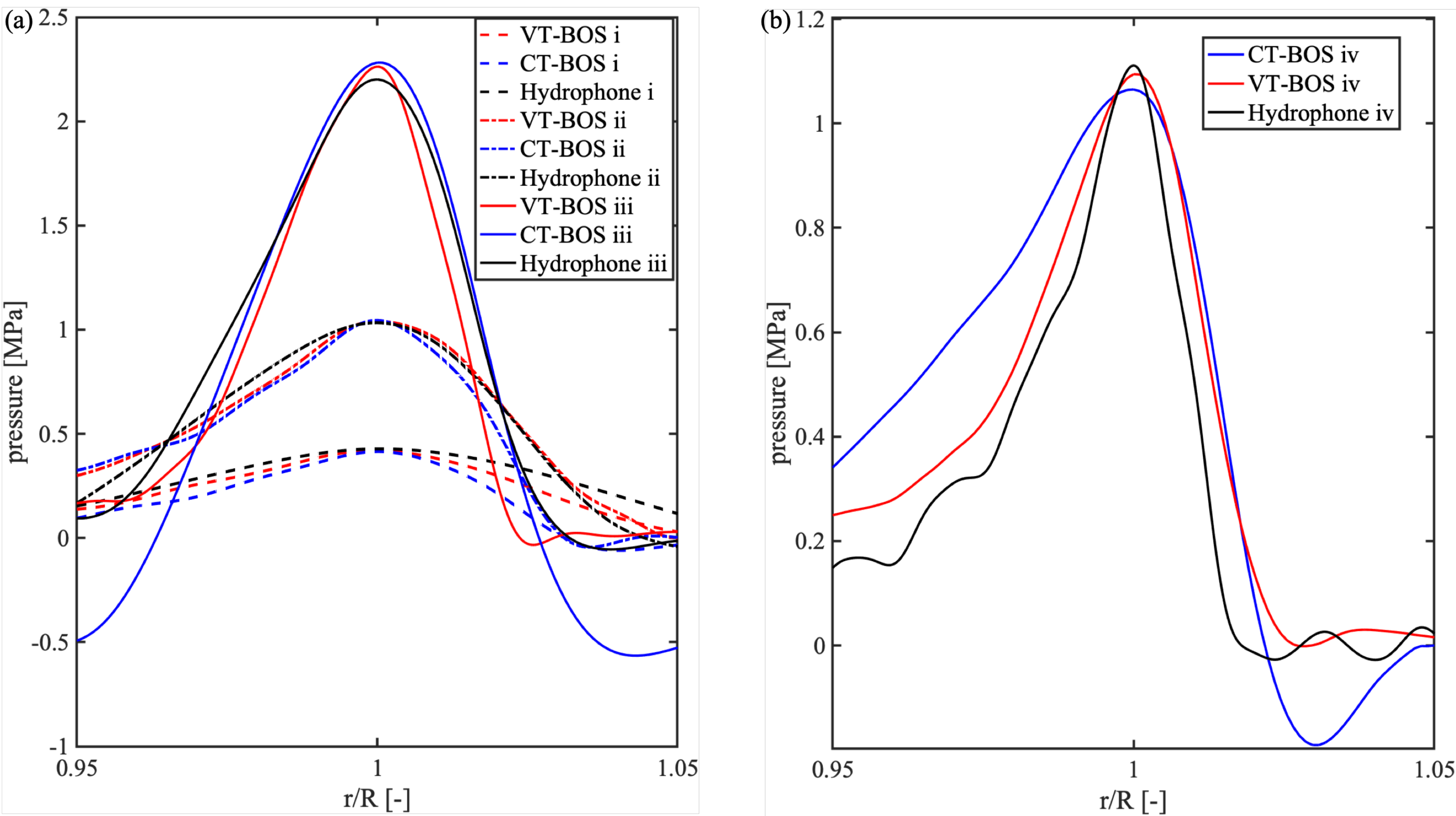}
\caption{Pressures of underwater shock wave obtained by BOS and hydrophone: (a)  datasets i--iii; (b)  dataset iv.}
\label{fig:pressure}
\end{figure}
The pressure waves from  VT-BOS, CT-BOS, and the hydrophone are shown in Fig.~\ref{fig:pressure}.
Note that the calculation criteria for the CT-BOS pressure calculation results will be discussed later in Sec. \ref{convergence}.

Fig.~\ref{fig:pressure}(a) shows that the VT-BOS pressure wave is closer to the hydrophone pressure wave than the CT-BOS pressure wave. 
Especially in the ranges  $0.95\le r/R\le0.98$ and $1.03\le r/R\le1.05$, the pressure wave from CT-BOS has relatively high and low values compared with that from the  hydrophone.
This trend is more pronounced in dataset iii. On the other hand, 
in Fig.~\ref{fig:pressure}(b), the pressure wave from VT-BOS is similar to that from the hydrophone in the range $0.95\le r/R\le1.05$.
This is because, unlike CT-BOS, VT-BOS does not involve the application of a differential operator, making its measurement accuracy less sensitive to noise.
Further verification is described in Sec.~\ref{sec:4_sp_syn}.

To quantitatively compare the accuracy of pressure calculation between VT-BOS and CT-BOS, the results for the ratio $S$ introduced in Sec.~\ref{sec:3_ratio} are shown in Table~\ref{tab:pre_diff}.
The ratio for VT-BOS  is lower than that for CT-BOS.
Furthermore, the error  between the maximum measured pressure and the maximum CT-BOS pressure is less than 4.15\%, as shown in Table~\ref{tab:err}.
On the other hand, the error  for VT-BOS is less than 2.78\%.
These results show that VT-BOS has higher accuracy in pressure calculation than CT-BOS.
In particular, VT-BOS tends to calculate pressure more accurately than CT-BOS in  ranges of  $r/R$ where the pressure gradient drops rapidly, such as $0.95\le r/R \le 0.98$ and $1.03 \le r/R \le 1.05$.

\begin{table}[H]
\centering
\caption{Ratio $S$ comparing  pressure measurements from CT-BOS and VT-BOS with hydrophone measurements.}
\begin{tabular}{l c c c c} 
Dataset& i  & ii& iii&iv\\ 
&  & & &\cite{hayasaka2016optical}\\ \hline \hline 
VT-BOS  & 1.30 & 1.14 & 1.18 & 1.27 \\ 
CT-BOS  & 1.47 & 1.17 & 1.34 & 1.70 \\
\end{tabular}
\label{tab:pre_diff}
\end{table}

\begin{table}[H]
\centering
\caption{Error [\%] in the pressure measurements using CT-BOS and VT-BOS.}
\begin{tabular}{l c c c c} 
Dataset& i  & ii& iii&iv\\ 
&  & & &\cite{hayasaka2016optical}\\ \hline \hline 
VT-BOS  & 1.47 & 0.66 & 2.78 & 1.51 \\
CT-BOS  & 3.11 & 1.13 & 3.66 & 4.15 \\ 
\end{tabular}
\label{tab:err}
\end{table}
\subsubsection{Convergence}
\label{convergence}
\begin{figure}[H]
\centering
\includegraphics[width=0.8\columnwidth]{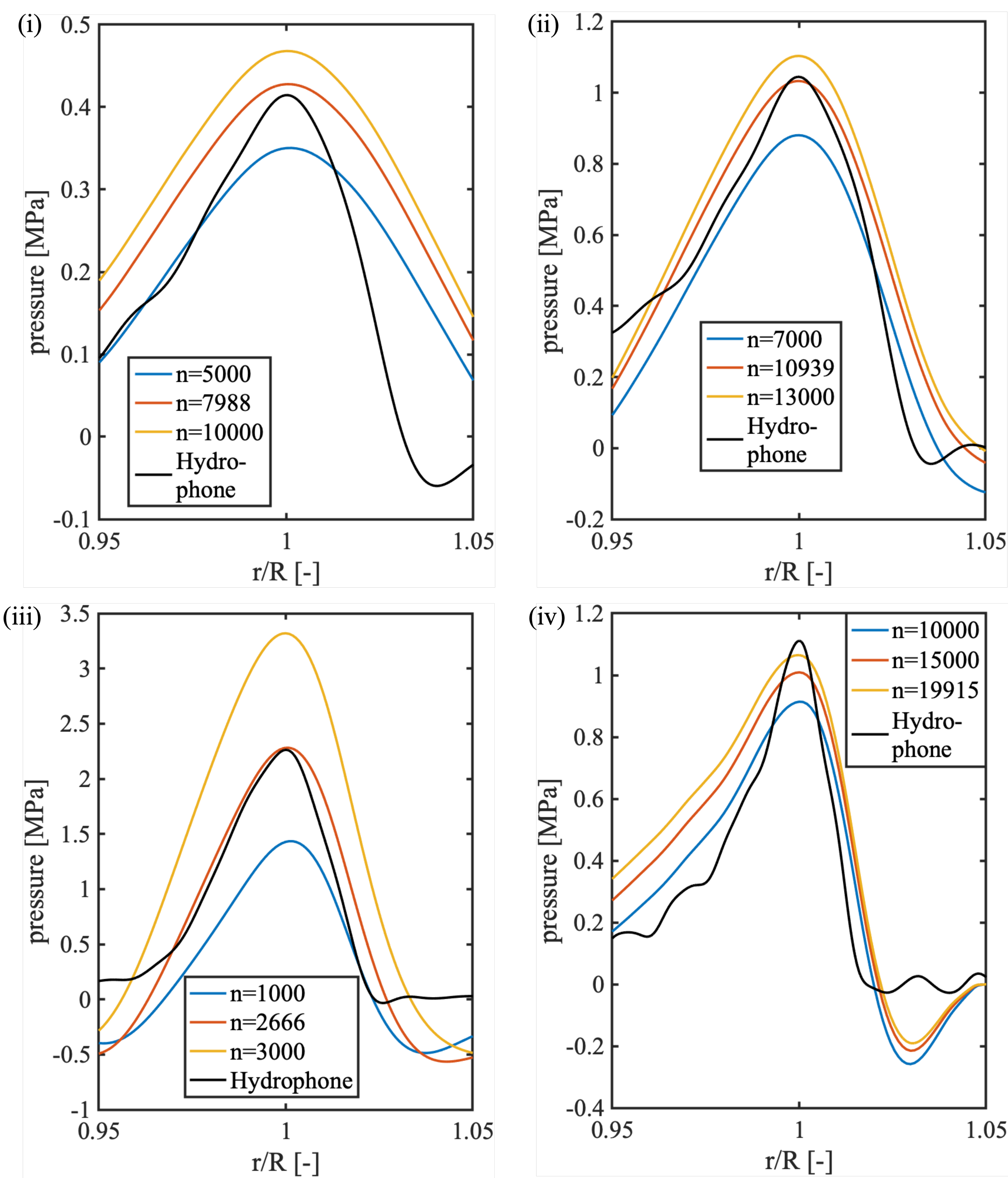}
\caption{Iteration results for a convergence value $E_0=1.0\times 10^{-8}$ for datasets i--iv. The experimental conditions of datasets i--iii are shown in Table~\ref{tab:exp_cond}, and dataset iv comprises the data from \cite{hayasaka2016optical}.}
\label{fig:convergence}
\end{figure}

\begin{figure}[H]
\centering
\includegraphics[width=0.8\columnwidth]{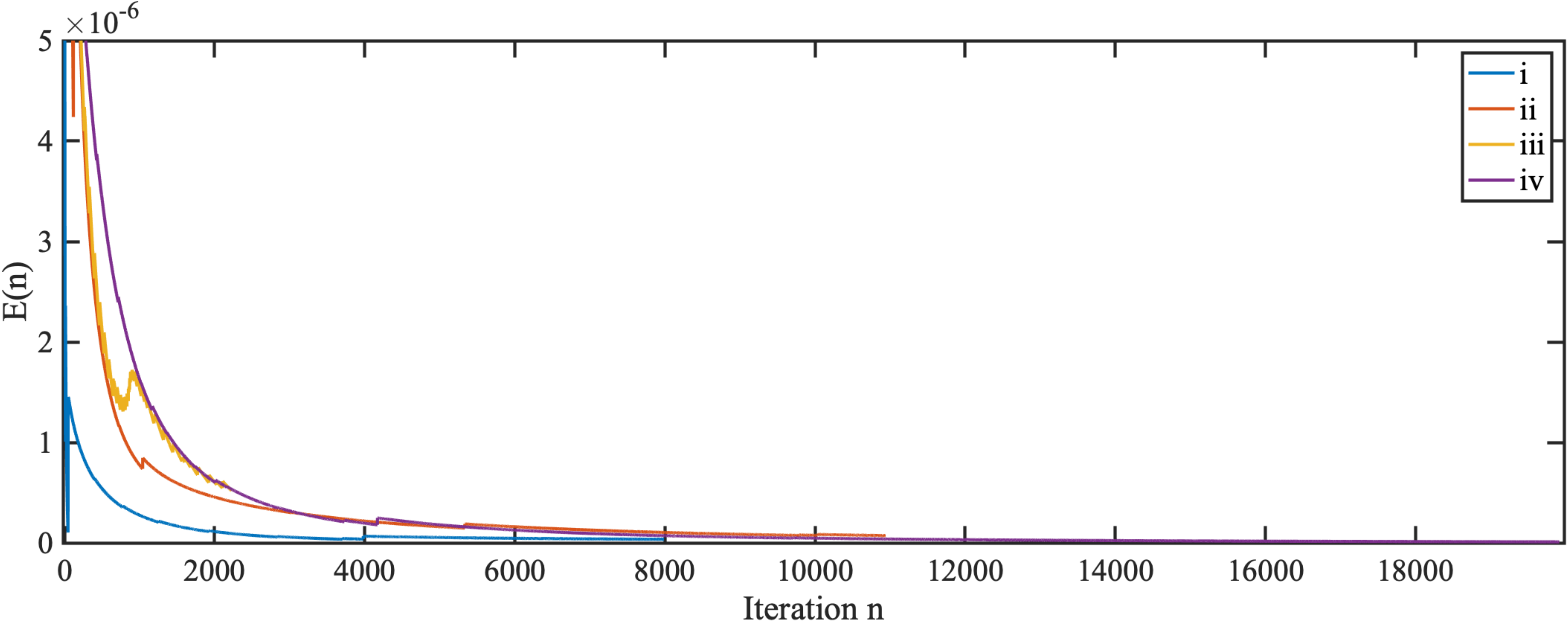}
 \caption{$E(n)$ of  datasets i--iv at each iteration step until the convergence condition~\ref{eq:convergence}  is satisfied. }
\label{fig:convergence_err}
\end{figure}

The convergence of the CT-BOS solution is discussed with reference to Figs.~\ref{fig:convergence} and \ref{fig:convergence_err} and Table~\ref{tab:convergence_E}. 
Fig.~\ref{fig:convergence} shows the CT-BOS iteration results for a convergence value $E_0 = 1.0\times 10^{-8}$\cite{hayasaka2016optical}.
However, for datasets i--iii, when the convergence condition~\ref{eq:convergence} is satisfied, the pressure waveforms exhibit larger values compared with the hydrophone pressure waveforms.
On the other hand, for  dataset iv, the  convergence condition~\ref{eq:convergence} is satisfied at iteration step $n=19\,915$, and the pressure waveforms from the hydrophone and CT-BOS are similar in shape.
It is not feasible to plot the  waveforms for all iteration steps in Fig.~\ref{fig:convergence}.
Therefore, for  datasets i--iii, we have plotted the results for the minimum value of $n$ such that  the maximum  pressure from CT-BOS is greater than or equal to the maximum pressure from the hydrophone, together with the results for two other values of $n$ respectively larger and smaller than this value.
For dataset iv, we have arbitrarily chosen a number of iteration steps $n=19\,915$ such that  the convergence condition~\ref{eq:convergence} is satisfied, together with  two other values of $n$ less than 19\,915.
In all data in Fig.~\ref{fig:convergence}, the pressure from CT-BOS  increases with increasing number of iteration steps.
If the maximum pressure from CT-BOS is close to the maximum pressure measured by the hydrophone, then number of iteration steps required for convergence depends on the experimental conditions.

 Table~\ref{tab:convergence_E} shows the convergence values $E_0$ when the maximum pressure from CT-BOS is closest to the maximum measured pressure  (the red curves in Figs.~\ref{fig:convergence}(i)--\ref{fig:convergence}(iii)).
It can be seen that $E_0$  is not the same for  the four datasets (i)--(iv).

Fig.~\ref{fig:convergence_err} shows  $E(n)$  for each number of iteration steps until the convergence condition~\ref{eq:convergence} is satisfied.
For datasets i, ii, and iv, $E(n)$ changes only by an amount of the order of $10^{-8}$ when the number of iteration steps exceeds 4000, while for dataset iii, $E(n)$ changes only by an amount of the order of $10^{-7}$ when the number of steps exceeds 1800.
Therefore, the CT-BOS iterations converge, but if the peak of the measured pressure and that of the pressure from CT-BOS  are to coincide, it may be necessary to take into account not only the convergence condition~\ref{eq:convergence}, but  other conditions as well.

\begin{table}[H]
\centering
\caption{Convergence value $E_0$ of each dataset.}
\begin{tabular}{l c c c c} 
Dataset& i  & ii& iii&iv\\ 
& & & &\cite{hayasaka2016optical} \\ \hline \hline
$E_0$ & $3.59\times 10^{-8}$ & $7.25\times 10^{-8}$ & $5.19\times 10^{-7}$  &$1.0\times 10^{-8}$ \\ 
\end{tabular}
\label{tab:convergence_E}
\end{table}

\subsection{Relationship to spatial resolution}
The relationship between the accuracy of pressure calculation and the spatial resolution of VT-BOS and CT-BOS is compared using the ratio $S$.
The method of varying the spatial resolution was described in Sec.~\ref{sec:3_sp}.
In Sec.~\ref{sec:4_sp_result}, we investigate the relationship between the spatial resolution and the measurement accuracy of the BOS technique using the four sets of experimental data given in Table~\ref{tab:exp_cond}.
In Sec.~\ref{sec:4_sp_syn}, we use synthetic data to investigate the reason for the strong dependence of CT-BOS measurement accuracy on spatial resolution, in contrast to that of VT-BOS.

\subsubsection{\label{sec:4_sp_result}Comparison  using experimental data}
\begin{figure}[H]
\centering
\includegraphics[width=0.7\columnwidth]{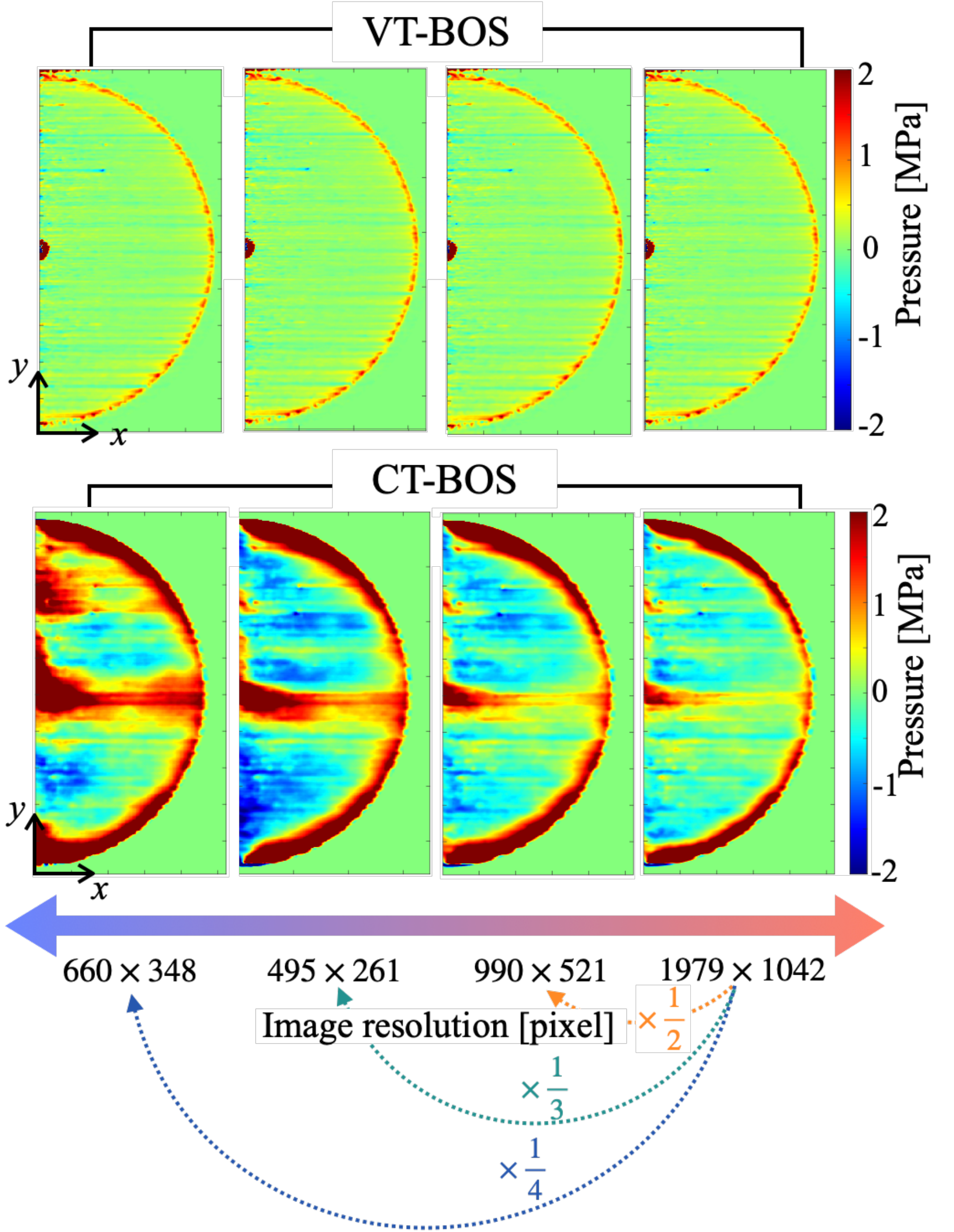}
\caption{Pressure fields of underwater shock wave from dataset iv measured by VT-BOS (top) and CT-BOS (bottom) at different image resolutions.
}
\label{fig:sp_field}
\end{figure}

Fig.~\ref{fig:sp_field} shows the pressure field calculated by the BOS technique for the data of \cite{hayasaka2016optical}.
The upper and lower diagrams show the pressure fields of the shock wave calculated by VT-BOS and CT-BOS, respectively,
from the displacement fields corresponding to  spatial resolution factors $\eta = 1$, 1/2, 1/3, and 1/4 from  right to left.
For VT-BOS, the pressure field remains almost unchanged as the spatial resolution decreases.
On the other hand, for CT-BOS, when the spatial resolution of the pressure field decreases, an increase in pressure  is observed at the center of the shock wave near the $y$ axis, which is unrelated to the shock wave and should not occur.

\begin{figure}[H]
\begin{center}
\centering
\includegraphics[width=1.0\columnwidth]{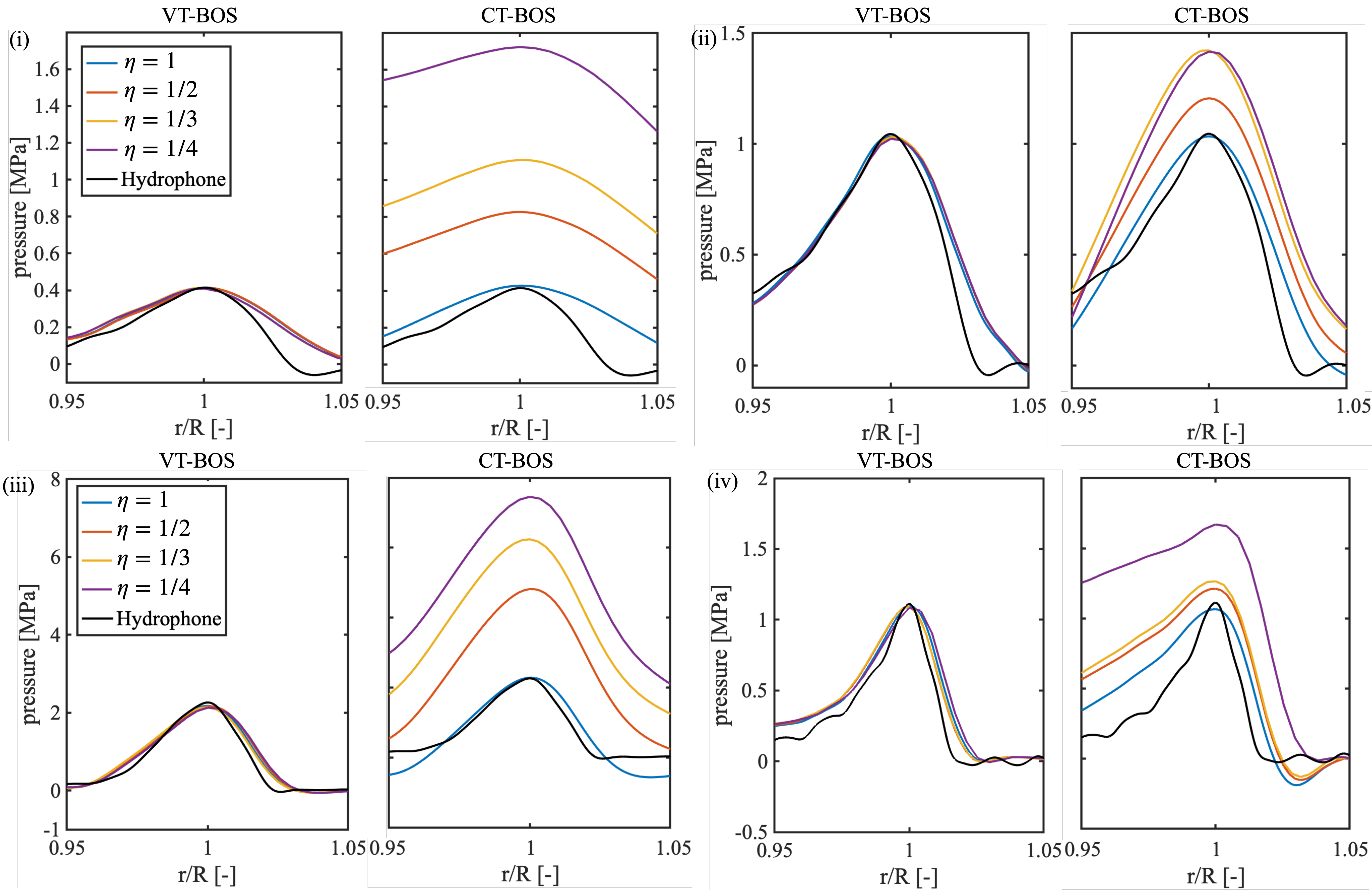}
\caption{Pressure measured by VT-BOS (left) and CT-BOS (right) at different spatial resolutions. (i)--(iv) show results from datasets i--iv, respectively.
}
\label{fig:sp_line}
\end{center}
\end{figure}

In Fig.~\ref{fig:sp_line}, the pressures from VT-BOS and CT-BOS are compared for each dataset.
In Figs.~\ref{fig:sp_line}(i)--\ref{fig:sp_line}(iv), the VT-BOS pressure waves remain almost unchanged with changes in spatial resolution.
The  pressure waves measured by the hydrophone and those from VT-BOS  are almost identical.
In Fig.~\ref{fig:sp_line}(i), for CT-BOS, as  the spatial resolution factor $\eta$ is decreased in the range from 1 to 1/4, the maximum pressure  increases about four times, and the measured pressure at each decrease in spatial resolution no longer coincides with the CT-BOS pressure.
In Fig.~\ref{fig:sp_line}(ii), the maximum pressure  from CT-BOS increases about 1.4 times as $\eta$ is decreased from 1 to 1/4.
As the spatial resolution decreases, the CT-BOS pressure no longer coincides with the measured pressure.
In Fig.~\ref{fig:sp_line}(iii), the  maximum pressure from CT-BOS increases by a factor of about three as  $\eta$ decreases from 1 to 1/4.
In Fig.~\ref{fig:sp_line}(iv), the  maximum pressure from CT-BOS increases about 1.5 times as $\eta$  decreases  from 1 to 1/4.
Thus, in all cases, at each decrease in spatial resolution, the measured pressure no longer coincides with the CT-BOS pressure.

\begin{figure}[H]
\centering
\includegraphics[width=0.7\columnwidth]{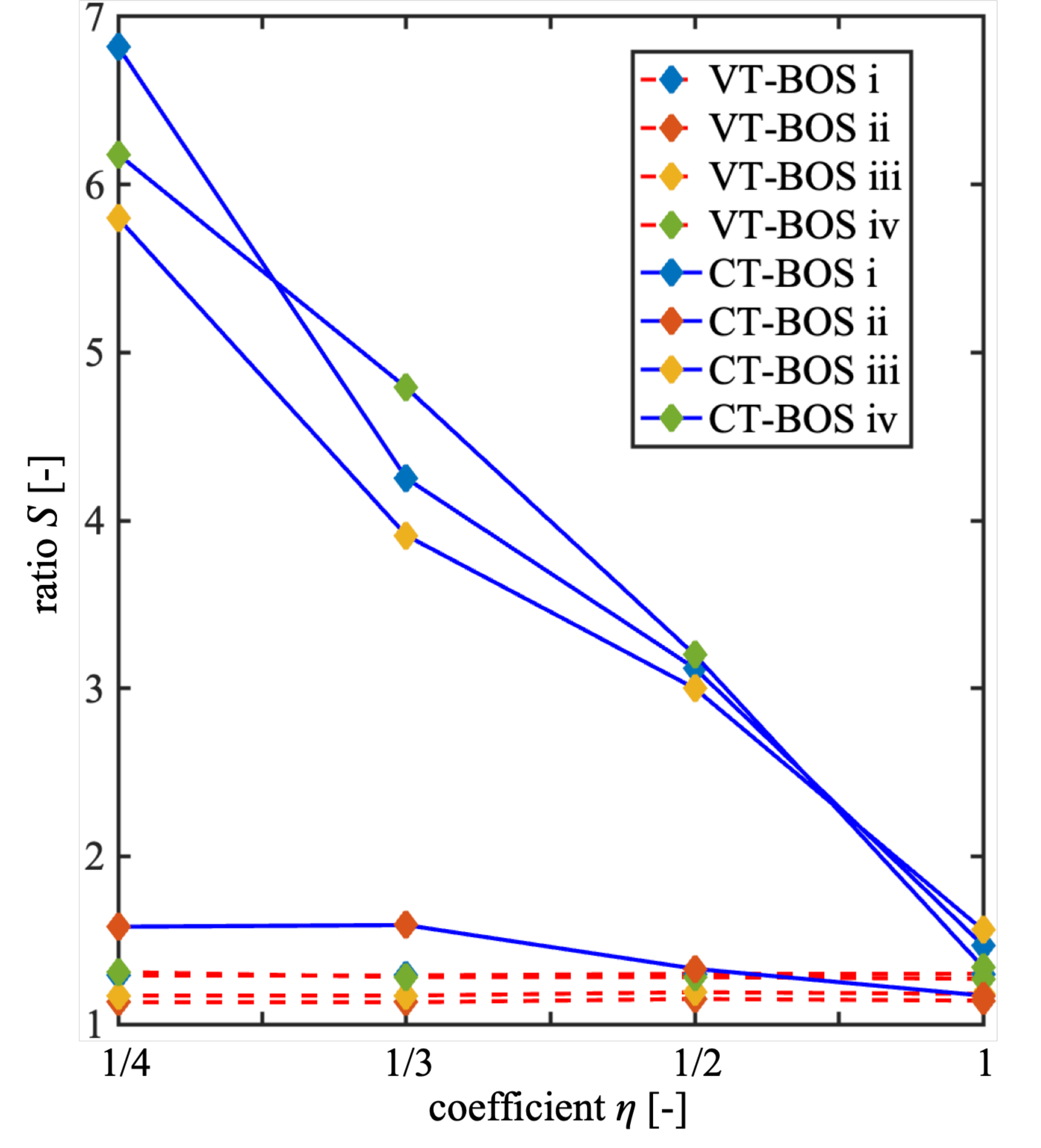}
\caption{Ratio $S$ at different spatial resolutions (represented by the factor $\eta$) for datasets i--iv. }
\label{fig:sp_ratio}
\end{figure}

Fig.~\ref{fig:sp_ratio} shows the ratio $S$, calculated as in  Sec.~\ref{sec:3_ratio}, for CT-BOS and VT-BOS as the spatial resolution factor $\eta$ is reduced in the range from 1 to 1/4.
It can be seen that $S$ increases with decreasing spatial resolution for CT-BOS.
For example, for CT-BOS~i, as $\eta$ is reduced to 1/4, the ratio increases by a factor of approximately 5 compared with the case $\eta=1$.
Similarly, $S$ increases by  factors of 1.3, 4 and 3 for CT-BOS~ii, CT-BOS~iii, and  CT-BOS~iv, respectively.
On the other hand, for VT-BOS, $S$ remains almost constant for all dataset, regardless of the change in spatial resolution.
In summary, similar to the results in Fig.~\ref{fig:sp_line}, the accuracy of the VT-BOS pressure calculation is hardly affected by the resolution.

\begin{figure}[H]
\centering
\includegraphics[width=0.7\columnwidth]{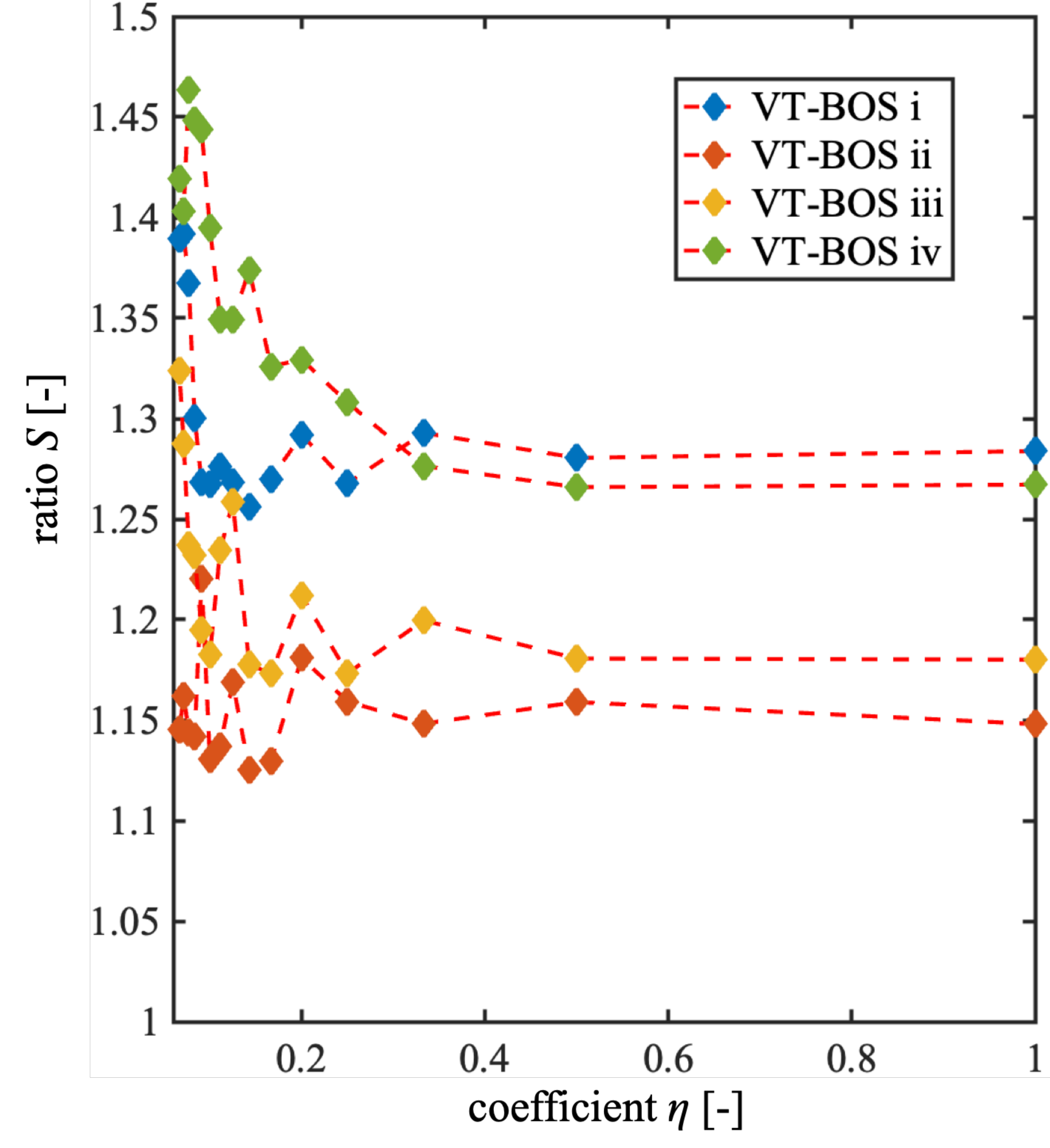}
\caption{Ratio $S$ as the spatial resolution factor  $\eta$ varies from 1 to 1/15 for datasets i--iv.}
\label{fig:sp_ratio_VT}
\end{figure}

To further investigate the effect of a reduction in spatial resolution  on the results from VT-BOS, the ratio $S$ is shown in Fig.~\ref{fig:sp_ratio_VT} as the spatial resolution factor $\eta$ is reduced from 1 to 1/15.
It can be seen that $S$ changes significantly  as $\eta$ decreases below 0.2. 
For example, for VT-BOS~i, $S=1.28$ for $\eta =1$, while $S=1.38$ for $\eta =1/15$, an increase of 0.1.
For VT-BOS~ii, $S=1.15$ for both $\eta =1$ and $1/15$.
For VT-BOS~iii and VT-BOS~iv, $S$ is increased by 0.14 and 0.15, respectively, for $\eta =1/15$  compared with $\eta =1$.
As can be seen from Fig.~\ref{fig:sp_ratio}, for CT-BOS, $S$ increases by factors of 4--5 as $\eta$  decreases from 1 to 1/4, whereas 
for VT-BOS, even when $\eta$ decreases below 0.2, although $S$ increases slightly,  this change can be regarded as almost negligible.

Therefore, compared with CT-BOS, the accuracy of pressure calculation by VT-BOS is nearly  independent of spatial resolution, and VT-BOS is able to give good results for the pressure field even when the spatial resolution is decreased.
In this context, the difference between the methods is the calculation procedure.
Differentiation operations  are performed twice in  CT-BOS (see  Fig.~\ref{fig:bos}), while   VT-BOS does not involve any differentiation at all.
Therefore, the dependence of CT-BOS measurement accuracy on resolution might be attributable  to the amplification of noise by the differentiation  operation.
This  will be  investigated further in the next subsection.


\subsubsection{\label{sec:4_sp_syn}Comparison of results using synthetic data}
To investigate in greater detail the cause of the difference between CT-BOS, whose measurement accuracy depends on spatial resolution, and VT-BOS, whose accuracy does not, we examine whether the presence of differentiation operations in the calculation procedure does indeed cause noise amplification.
The experimental data contain noise not only in the displacement field measured by the BOS technique, but also in the pressure measured by the hydrophone.
Therefore, when comparing   datasets, it is difficult to determine how much of the increased noise is due to amplification by  differentiation operations in  the BOS technique and how much is due to the noise in the hydrophone pressure.
The use of synthetic data allows the displacement field to be subjected to controllable noise, eliminating factors other than the displacement field's own noise.

\paragraph{Synthetic data}

\begin{figure}[H]
\centering
\includegraphics[width=1.0\columnwidth]{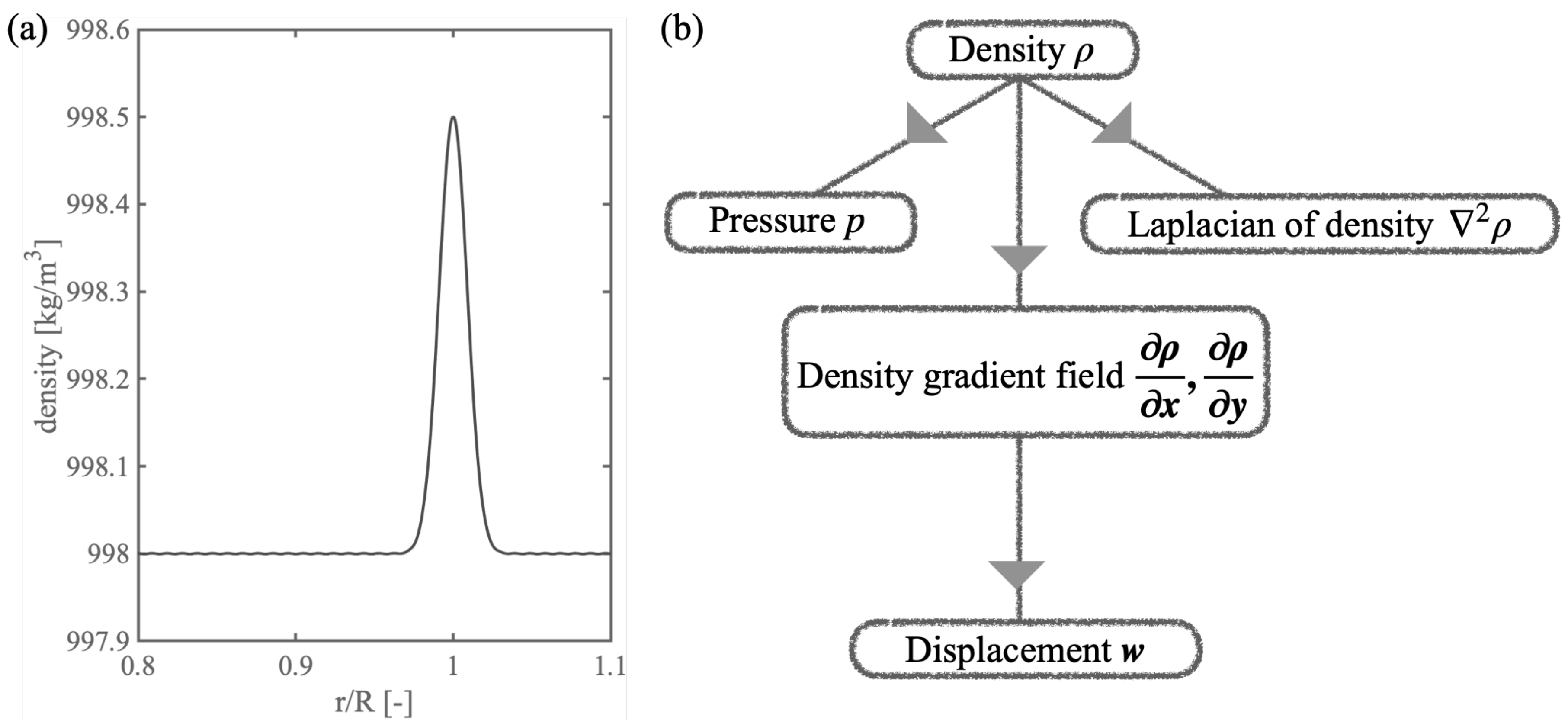}
\caption{(a) Synthetic $\rho (r)$ (Eq.~\ref{eq:num_rho}). (b) Synthetic data that can be generated from the density $\rho (r)$.}
\label{fig:num_rho}
\end{figure}

The synthetic data are provided by the following expression simulating the density field of underwater shock waves:
\begin{equation}
\rho(r) = \rho_{0} + 
\xi \exp \!\left[-\frac{(r-r_c)^2}{\gamma} \right] 
+ \varepsilon \sin ( \zeta r ),
\label{eq:num_rho}
\end{equation}
where $\rho_{0}$ is the density of the surrounding fluid, the exponential term represents the  change in density of the measurement target, and the sinusoidal term represents  noise.
Here, $r$  is the radial coordinate in a polar coordinate system $(r,\theta)$, as in Figs.~\ref{fig:recon_vr1}(b) and \ref{fig:recon_vr1}(c).
Eq.~\ref{eq:num_rho} represents a density field that does not depend on $\theta$ and is distributed in the $xz$ section centered at the origin, as shown in Fig.~\ref{fig:recon_vr1}.
Fig.~\ref{fig:num_rho}(a) is a plot of $\rho(r)$, and 
Fig.~\ref{fig:num_rho}(b) shows the quantities calculated using this density.
The density gradient field $\partial \rho / \partial r$  is  distributed on the $xz$ section (Fig.~\ref{fig:recon_vr1}) in the same way as $\rho (r)$.
The displacement $u$ is calculated from Eq.~\ref{eq:u-rho} by integrating $\partial \rho / \partial r$ along the $z$ direction.
When using the synthetic data from Eq.~\ref{eq:num_rho}), in order to obtain a  density field similar to that of an underwater shock wave, the following parameter values are adopted: $r_c = 2.5\times 10^{-3}$~m, $\xi=0.5$, $\gamma=1\times 10^{-9}$~m, $\varepsilon=5.0\times 10^{-4} \pi$/m , and $\zeta=2.7\times 10^{4}$.


The displacement $\bm{w}$ in Fig.~\ref{fig:bos} can be obtained by the above procedure.
In this subsection, we investigate the effects of the calculations of first and second derivatives in CT-BOS on the spatial resolution.
The spatial resolution of the  displacements generated using the synthetic data is reduced as in  Sec.~\ref{sec:4_sp_result}.
CT-BOS is applied to these synthetic displacements to investigate the effect of spatial resolution on the divergence  $\nabla\cdot\bm{w}$ of the displacement (i.e., after one differentiation operation).
However, as can be seen from Fig.~\ref{fig:num_rho}(b), the synthetic $\nabla\cdot\bm{w}$ cannot be calculated from the density $\rho$, and so the divergence $\nabla\cdot\bm{w}$ from CT-BOS and the synthetic $\nabla\cdot\bm{w}$ are not compared. 
Next, the Laplacian $\nabla^2 \rho$ of the density gradient from the CT-BOS calculation is  compared with the synthetic $\nabla^2 \rho$ calculated from the density $\rho(r)$ to examine the effect of reducing the spatial resolution when two differentiation operations have been performed.
Finally, we compare the pressure calculated by CT-BOS with the synthetic $p$ and discuss the impact of reduced spatial resolution.

\paragraph{Comparison with synthetic data}

\begin{figure}[H]
\centering
\includegraphics[width=0.8\columnwidth]{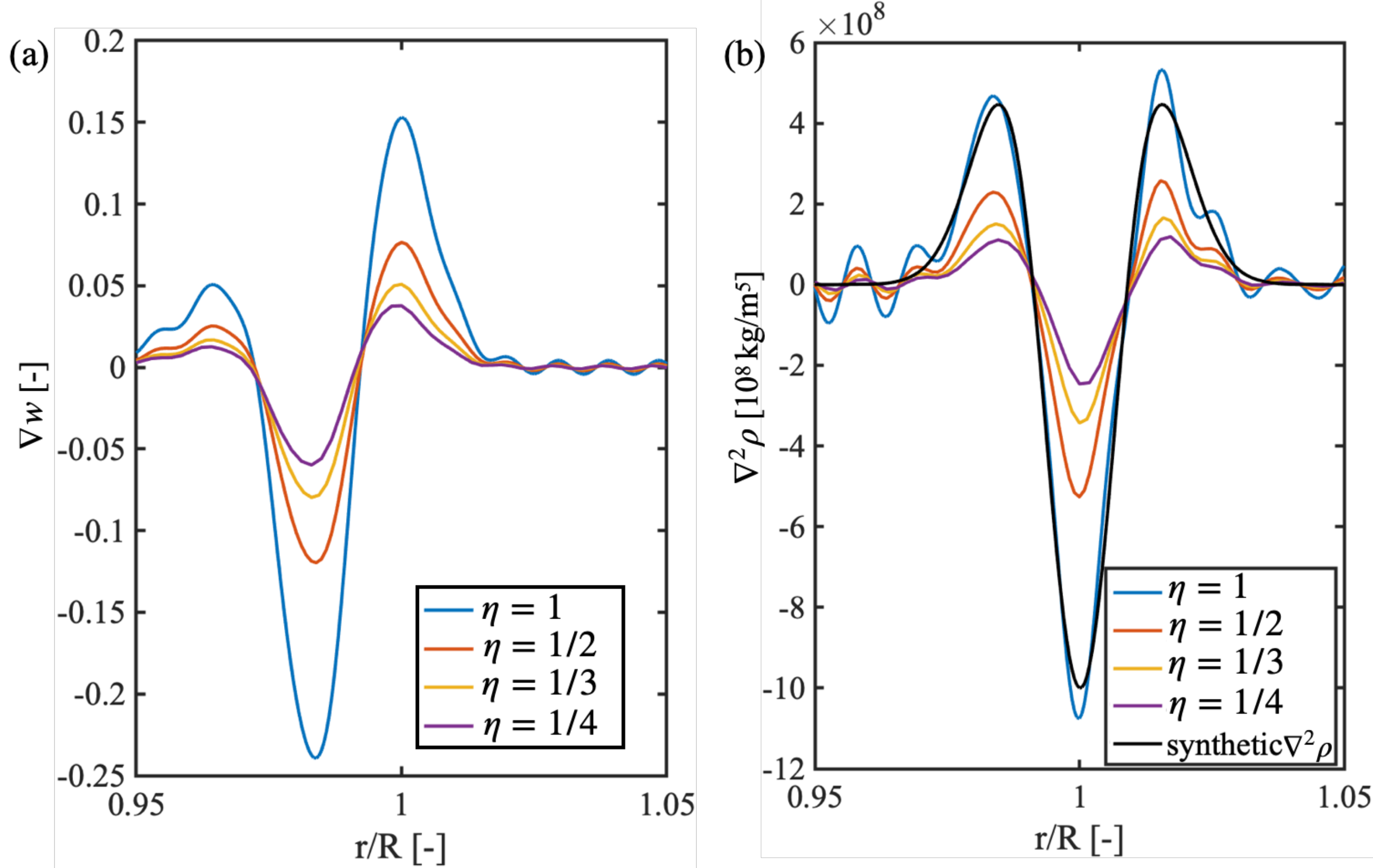}
\caption{(a) Divergence of displacement $\nabla \cdot \bm{w}$, (b)  Laplacian of density $\nabla^2 \rho$.}
\label{fig:num_u_rho}
\end{figure}

The CT-BOS calculation is applied to the synthetic data displacement to investigate the effect of spatial resolution on the derivative calculation.
Fig.~\ref{fig:num_u_rho} show the  relationships between the spatial resolution and the divergence  of the displacement (i.e., a first derivative) and the Laplacian of the density (i.e., a second derivative).
From Fig.~\ref{fig:num_u_rho}(a), it can be seen that the divergence  $\nabla\cdot\bm{w}$  for reduced  spatial resolution factors $\eta=1/2$, 1/3, and 1/4 is lower than that for $\eta=1$.
In particular, at $0.95 \le r/R \le 1.02$, the larger the value of $\eta$ (the greater the spatial resolution), the larger is the divergence  of the displacement $\bm{w}$.
The accuracy of the first derivative decreases in proportion to the decrease in spatial resolution.

From Fig.~\ref{fig:num_u_rho}(b), it can be seen that the difference between $\nabla^2 \rho$ from CT-BOS and that from the synthetic data increases as the spatial resolution decreases.
For  a spatial resolution factor $\eta=1$, the synthetic $\nabla^2\rho$ and the CT-BOS $\nabla^2\rho$ are almost identical, but
 as  $\eta$ decreases from 1/2 to 1/4, the magnitude of the CT-BOS $\nabla^2\rho$  decreases compared with that of the synthetic $\nabla^2\rho$.
In CT-BOS, $\nabla^2\rho$ is not calculated by differentiation operations, but by performing a scalar field 3D reconstruction, and 
the accuracy of this reconstruction is decreased when the spatial resolution is reduced.

The above results indicate that when CT-BOS is applied to a displacement field containing noise, the calculation accuracy decreases as the spatial resolution decreases in the calculation procedure for the first  derivative (taking the divergence of the displacement) and second derivative (obtaining the Laplacian of the density field by 3D reconstruction of a scalar field).
Unlike CT-BOS, VT-BOS does not involve differential operations and therefore  is not subject to degradation in measurement accuracy when the spatial resolution is reduced.

\subsection{Computational  cost}
The  computational costs incurred in calculating the pressure field from the displacement field using CT-BOS and VT-BOS, respectively, are now compared. 
The lower the  CPU time, the lower is the computational cost.

\begin{figure}[H]
\centering
\includegraphics[width=0.7\columnwidth]{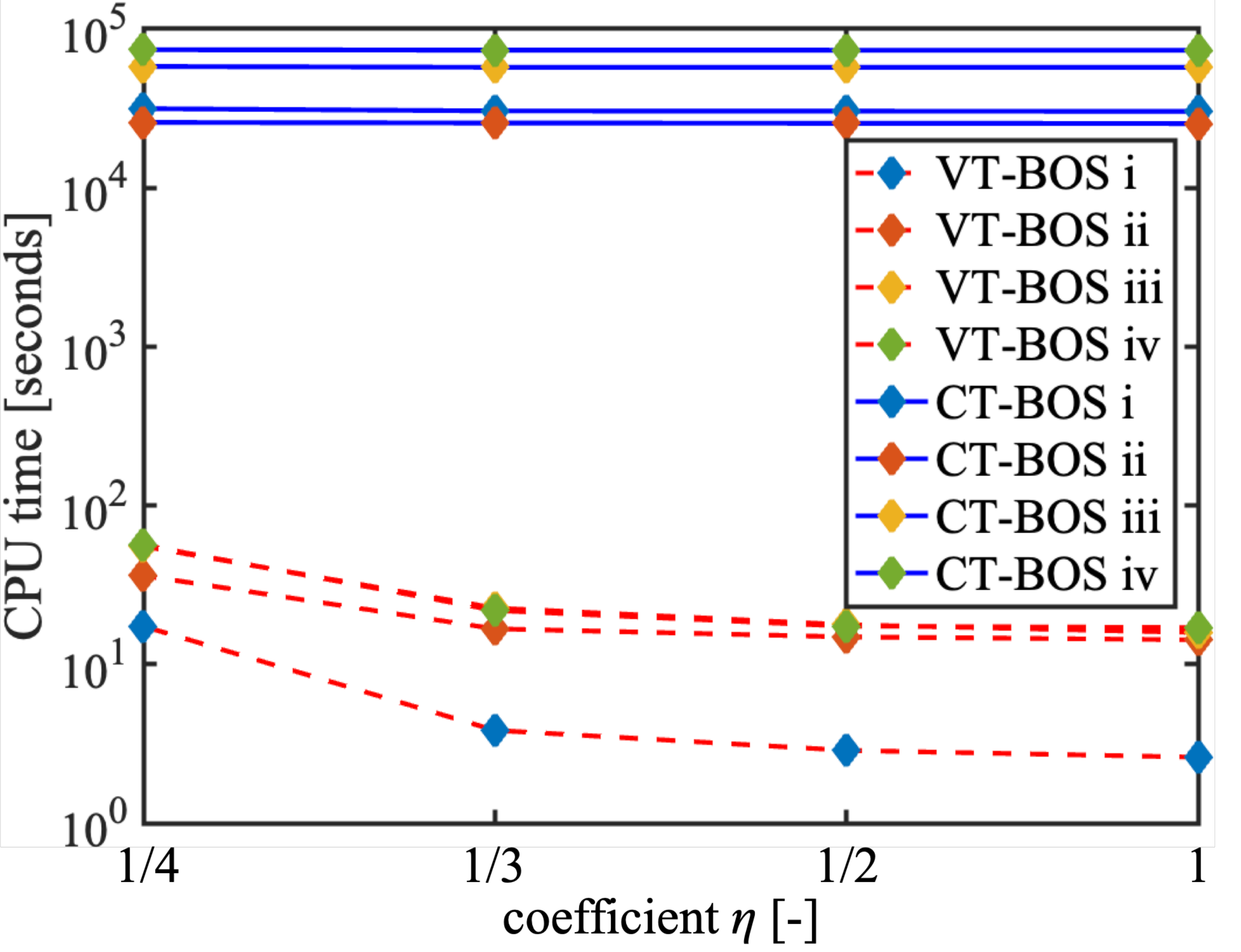}
\caption{CPU times of CT-BOS and VT-BOS at different spatial resolutions for datasets i--iv.
}
\label{fig:time}
\end{figure}

Fig.~\ref{fig:time} shows the CPU times of both techniques at different spatial resolutions.
For dataset i, the difference in CPU times  between CT-BOS and VT-BOS is about 30\,000~s at any spatial resolution.
At a spatial resolution factor  $\eta=1$, the CPU time of CT-BOS is 1800 times that of VT-BOS.
For dataset ii, the difference in CPU times is about 25\,000~s at all spatial resolutions.
At $\eta=1$, the CPU time of CT-BOS is 710 times that of VT-BOS.
For dataset iii, the difference in CPU times is approximately 58\,000~s at all spatial resolutions.
At $\eta=1$, the CPU time of CT-BOS is 1050 times that of VT-BOS.
For dataset iv, the difference in CPU times  is about 74\,000~s at all spatial resolutions.
At $\eta=1$, the CPU time of CT-BOS is 1300 times that of VT-BOS.
Thus, under all experimental conditions, the computational cost of VT-BOS is significantly lower than that of CT-BOS.

With VT-BOS, the computational cost also decreases as the spatial resolution decreases under all experimental conditions.
For example, for dataset i, there is a 15~s difference in CPU times between the case $\eta=$1 and the case $\eta=$1/4, and for datasets ii and iii, the difference in CPU times between the two cases is about 22--39~s.
With CT-BOS, the computational cost is nearly constant under all experimental conditions, independent of spatial resolution.

Thus, VT-BOS has an overwhelmingly lower computational cost than CT-BOS, because, rather than 3D scalar field reconstruction, VT-BOS uses a matrix equation (Eq.~\ref{eq:vr_bos}) for VT, which does not require iterative calculations.
Specifically, in VT, Eq.~\ref{eq:vr_bos} can be computed by computing a square matrix of constant $\bm{A}$ once.
In VT-BOS, the lengths of the rows and columns of the constant matrix $\bm{A}$ decrease as the spatial resolution decreases, and so the computational cost decreases as the spatial resolution decreases.

\begin{figure}[H]
\centering
\includegraphics[width=0.7\columnwidth]{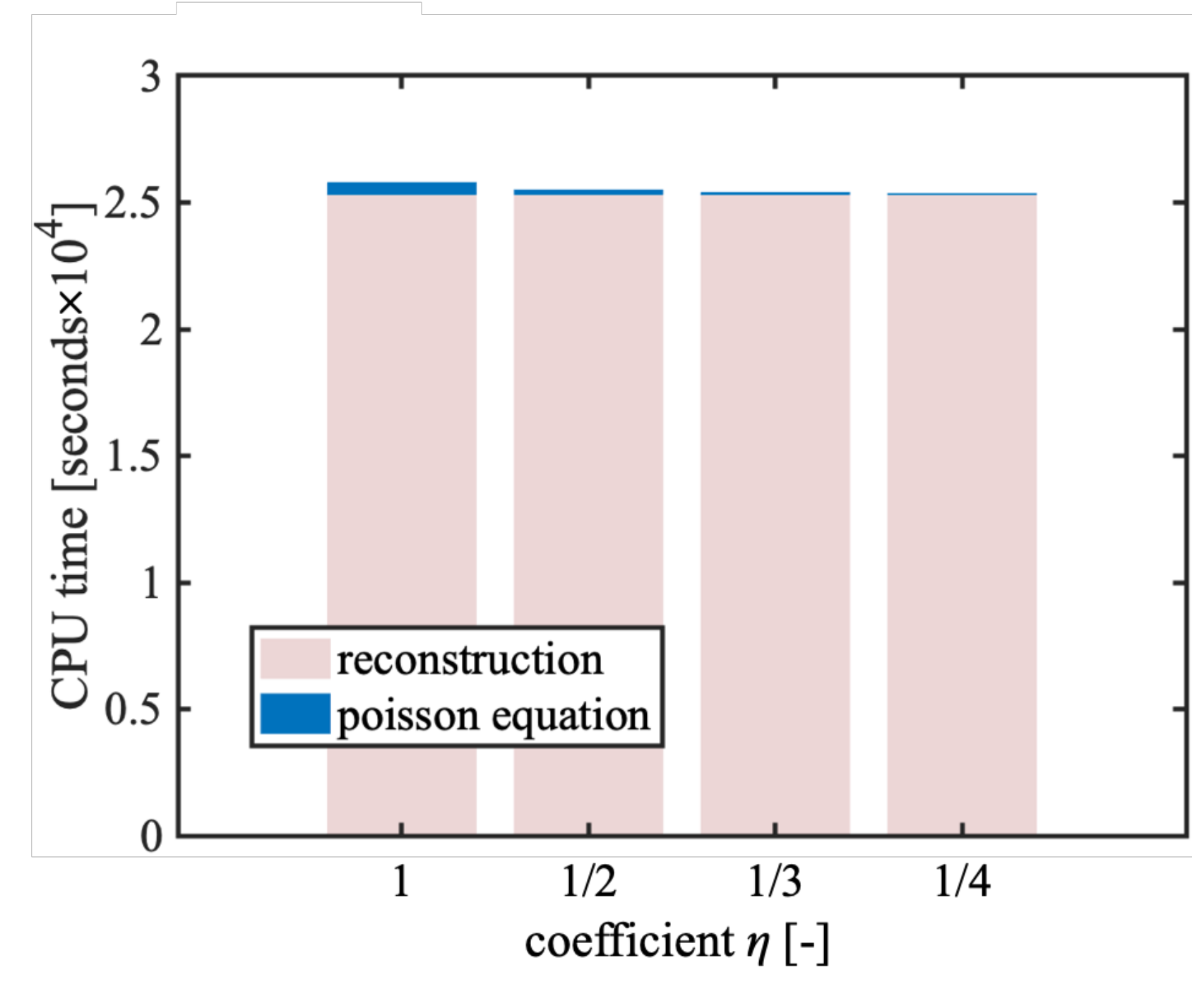}
\caption{CPU times of CT-BOS at different spatial resolutions for dataset i. The pink bars represent the CPU time for 3D reconstruction of the scalar field from the displacement, and the blue bars represent the CPU time for calculation of the pressure field from the Laplacian of the density.}
\label{fig:time_bar}
\end{figure}

 We now discuss the reasons why the computational cost of CT-BOS is independent of the resolution, referring to Fig.~\ref{fig:time_bar}, which  shows   CPU times  at different spatial resolutions for dataset i.  
These CPU times are for 3D reconstruction of the scalar field from the displacement field and for calculation of the pressure from the Laplacian of the density via the Poisson equation. It can be seen that  
more than 90\% of the total computational cost incurred by CT-BOS is for 3D reconstruction of the scalar field and that this cost remains  constant regardless of the resolution. On the other hand,  the comparatively small cost associated with the Poisson equation does decrease with decreasing resolution.


From Figs.~\ref{fig:time} and \ref{fig:time_bar} reveal that the computational cost of CT-BOS is almost independent of the spatial resolution, whereas that of VT-BOS does vary with resolution, depending on the experimental conditions. 

\section{\label{sec:5}Conclusions}
The purpose of this study was to measure the noncontact pressure field of a laser-induced underwater shock wave using the background-oriented schlieren (BOS) technique with a new vector tomography (VT) technique.
In VT, the  distribution of a vector field (in this case the refractive index gradient of an underwater shock wave) is reconstructed from the projected value of a vector field (in this case the displacement vector), which is integrated along one direction.
Conventional VT can only perform reconstruction if the divergence  of the reconstructed distribution is zero.
Therefore, it has been difficult to apply VT to  measurement targets (e.g., underwater shock waves) that do not satisfy this condition.
In this paper, we have assumed that the reconstructed distribution is axisymmetric and has only a radial component, and we have constructed an approximate matrix equation for VT that relates the projected values and the reconstructed distribution.

To evaluate VT-BOS, experimental data and data from a previous study \cite{hayasaka2016optical} have been used to compare it with computed tomography (CT)-BOS from three aspects: accuracy and convergence of pressure calculation, dependence on spatial resolution, and computational cost.
The results can be summarized as follows:
\begin{enumerate}
\item The accuracy of pressure calculation is higher with VT-BOS than with CT-BOS.
This is especially true in the range where the gradient of the measurement target changes rapidly.
In the case of VT-BOS, the use of the Poisson equation is avoided and the pressure can be calculated uniquely using a matrix equation.
\item In comparison with the experimental data, the accuracy of the CT-BOS pressure calculation decreases as the spatial resolution decreases, whereas the accuracy of the VT-BOS pressure calculation is almost independent of  spatial resolution within the conditions tested in this research.
Results obtained using synthetic data show that the dependence of CT-BOS measurement accuracy on spatial resolution is due to the use of differentiation operations  and 3D scalar field  reconstruction. 
VT-BOS does not involve either differentiation operations or 3D scalar field  reconstruction, and instead uses linear interpolation (Eq.~\ref{eq:vr_interp}), as a consequence of which the accuracy of pressure calculation is virtually unaffected by the spatial resolution within the conditions tested in this research.
\item Compared with CT-BOS, VT-BOS incurs a far lower computational cost.
This is because VT-BOS does not require iterative calculations.
Also, as a consequence of point~2, VT-BOS has the advantage of further reducing the computational load by reducing the requirements on spatial resolution, since its accuracy remains almost constant even when the spatial resolution is reduced.
\end{enumerate}

For targets that can be assumed to be axisymmetric, such as laser-induced underwater shock waves, the BOS method with the new VT technique is more suitable  for calculating the pressure field than the BOS method with 3D scalar field reconstruction.  
Unlike conventional VT, the technique proposed in this paper does not require that the divergence  of the reconstructed distribution be zero.
Therefore, it can be used not only for  measurements in fluids, but also for measurements of axisymmetric  objects with only radial components in electromagnetics, material mechanics, and other fields.


\bibliographystyle{ieeetr}
\bibliography{ref}
\end{document}